\documentclass[useAMS,usenatbib]{mnras}
\usepackage{graphicx}
\usepackage{txfonts}
\usepackage{color} 

\title[Environmental effects on galactic abundances]
      {On the influence of the environment on galactic chemical abundances} 
      
\author[Pilyugin et al.]
       {L.~S.~Pilyugin$^{1,2}$,
        E.~K.~Grebel$^{2}$,
        I.~A.~Zinchenko$^{1,2}$,
        Y.~A.~Nefedyev$^{3}$,
        L.~Mattsson$^{4}$    \\
       $^{1}$ Main Astronomical Observatory
             of National Academy of Sciences of Ukraine,
             27 Zabolotnogo str., 03680 Kiev, Ukraine \\   
       $^{2}$ Astronomisches Rechen-Institut, Zentrum f\"{u}r Astronomie 
             der Universit\"{a}t Heidelberg, 
             M\"{o}nchhofstr.\ 12--14, 69120 Heidelberg, Germany \\
       $^{3}$ Kazan Federal University, 18 Kremlyovskaya St., 420008, Kazan. Russian Federation \\
       $^{4}$ Nordita, KTH Royal Institute of Technology \& Stockholm University, Roslagstullsbacken 23, SE-106 91, Stockholm, Sweden
}
\date{Accepted 2016 November 1. Received 2016 October 29; in original form 2016 August 23}

\begin{document}

\maketitle

\begin{abstract} 
We examine the influence of the environment on the chemical abundances
of late-type galaxies with masses of $10^{9.1}$ M$_{\sun}$ --
$10^{11}$ M$_{\sun}$ using data from the Sloan Digital Sky Survey
(SDSS).  We find that the environmental influence on galactic chemical
abundances is strongest for galaxies with masses of $10^{9.1}$
M$_{\sun}$ to $10^{9.6}$ M$_{\sun}$. The galaxies in the densest
environments may exceed the average oxygen abundances by about
$\sim0.05$ dex (the median value of the overabundances for 101
galaxies in the densest environments) and show higher abundances in
nitrogen by about $\sim0.1$. The abundance excess decreases with
increasing galaxy mass and with decreasing environmental density.
Since only a small fraction of late-type galaxies is located in
high-density environments these galaxies do not have a significant
influence on the general X/H -- $M$ relation. The metallicity -- mass
relations for isolated galaxies and for galaxies with neighbours are
very similar.  The mean shift of non-isolated galaxies around the
metallicity -- mass relation traced by the isolated galaxies is less
than $\sim0.01$ dex for oxygen and less than $\sim0.02$ dex for
nitrogen.  The scatter in the galactic chemical abundances is large
for any number of neighbour galaxies (at any environmental density),
i.e., galaxies with both enhanced and reduced abundances can be found
at any environmental density.  This suggests that environmental
effects do not play a key role in evolution of late-type galaxies as
was also concluded in some of the previous studies. 
\end{abstract}


\section{Introduction}

Establishing the parameters that govern the (chemical) evolution of
galaxies is very important for understanding their formation and
evolution.  \citet{Lequeux1979} have shown that the metallicity in a
galaxy correlates with its mass.  The existence of a metallicity --
mass (or metallicity -- luminosity) relation at the present epoch and
at different redshifts has been confirmed in many investigations
\citep[][among many others]{Lequeux1979,Zaritsky1994, Garnett2002,
Grebel2003, Tremonti2004, Erb2006, Cowie2008, Maiolino2008,
Guseva2009, Thuan2010, PilyuginThuan2011, Pilyugin2013, Andrews2013,
Zahid2013, Maier2014, Steidel2014, Izotov2015}.  Nonetheless, the
scatter in the abundances X/H at a given galaxy mass suggests that
there may be additional parameters affecting galactic chemical
evolution.
  
The gas-phase metallicity of galaxies with higher present-day star
formation rates tends to be lower at a given stellar mass
\citep[][among
others]{Ellison2008,Zhang2009,LaraLopez2010,Mannucci2010,Bothwell2013}.
This can be easily understood from the following consideration. The
abundance excess in a galaxy indicates that that galaxy has a higher
astration level (and, consequently, a lower gas mass fraction) in
comparison to the average galaxy (with zero abundance excess) at that
stellar mass. Since the star formation rate decreases with decreasing
amount of gas
\citep[e.g.,][]{Kennicutt1998,Daddi2010,KennicuttEvans2012} the star
formation rate in galaxies with an abundance excess should be lower in
comparison to standard galaxies of given stellar mass. It is difficult
to consider the star formation rate as an independent parameter that
governs the chemical evolution of galaxies. In general, the star
formation history of a galaxy is defined by its mass. 

The distribution of galaxies in the universe forms a complex network
known as the cosmic web, comprising clusters, groups, filaments,
walls, and voids \citep[e.g.,][]{Zeldovich1982, deLapparent1986,
Bond1996, Alpaslan2014}.  There are also very large overdense
structures. The largest overdense structures in the nearby universe
(the Shapley supercluster or Shapley concentration, the Great
Attractor, and the Sloan Great Wall) include thousands of galaxies
\citep{Shapley1930,LyndenBell1988,Gott2005,Sheth2011}.  The dependence
of galaxy properties on environment has been considered in many papers.
There is a well-defined relation between the local density of galaxies
and the relative numbers of different morphological types.  This
morphology-density relation extends over five orders of magnitude in
density in the sense that the fraction of spiral and irregular
galaxies decreases with increasing density 
\citep[e.g.,][]{Dressler1980,Poggianti1999,vanderWel2008}.  On the one
hand, this implies that the environment affects the star formation
history of galaxies.  On the other hand, it was found that only the
fraction of the star-forming galaxies depends on the environment while
the star formation rates of the star-forming galaxies are similar in
the highest and lowest density environments
\citep{Balogh2004,Brough2013,Wijesinghe2012,Beygu2016}.

An influence of the environment on galaxy evolution should result in a
dependence of galactic metallicity on environment.  Many studies have
been devoted to the investigation of the dependence of galactic
chemical abundances on galactic environment \citep[][among many
others]{Shields1991, Vilchez1995, Skillman1996,Pilyugin2002,
Mouhcine2007, Ellison2009, Hughes2013, Peng2014, Pustilnik2013,
Nicholls2014, Darvish2015, Kacprzak2015, Kreckel2015, Shimakawa2015,
Valentino2015}, but no firm, unanimous conclusion has been reached.
In fact, recent studies arrived at contradictory results.
\citet{Kacprzak2015} found that there is no discernible difference
between the mass -- metallicity relation of field and cluster galaxies
to within 0.02 dex.  \citet{Shimakawa2015}, on the other hand,
established that cluster galaxies tend to be more chemically enriched
than their field counterparts.  In contrast, \citet{Valentino2015}
suggest that star-forming galaxies in clusters are more metal-poor
than their field counterparts. 
\citet{Gupta2016} investigated two galaxy clusters at redshifts 
$z$ $\sim$0.35. They found that the mass-metallicity relation for one cluster 
had an offest of 0.2 dex to higher metallicity compared to the local 
reference sample at a fixed mass. The median offsets in metallicity 
for galaxies in the other cluster is within the errors bars of the observations.

Our study of the dependence of galactic metallicity on its environment
is motivated by the following.  The present-day metallicity of a
star-forming galaxy is usually specified by the oxygen abundance in
the interstellar medium, because oxygen is the most abundant heavy
element in the interstellar medium, because its abundance can be
easily measured via emission lines when the gas is ionized, and
because the majority of the methods for abundance determinations were
developed for the measurement of oxygen abundances.  Recently, we
obtained a new calibration that provides estimates of the oxygen and
nitrogen abundances in star-forming regions with high precision
over the whole metallicity scale \citep{Pilyugin2016}.  
The mean difference between the strong-line based and $T_{e}$-based 
abundances is less than 0.05 dex for both oxygen and nitrogen. 
Since at 12 + log(O/H) $\ga$ 8.2, secondary
nitrogen becomes dominant and the nitrogen abundance increases at a
faster rate in comparison to the oxygen abundance
\citep{Pagel1997,Henry2000} the change in nitrogen abundance has a
larger amplitude than the one in oxygen abundance and, as a
consequence, may be easier to detect.  Our present investigation is
based on accurate oxygen and nitrogen abundances in a large sample of
SDSS galaxies.

We will use the following standard notations for the line intensities: \\
$R_2$  = $I_{\rm [O\,II] \lambda 3727+ \lambda 3729} /I_{{\rm H}\beta }$,  \\
$N_2$  = $I_{\rm [N\,II] \lambda 6548+ \lambda 6584} /I_{{\rm H}\beta }$,  \\
$S_2$  = $I_{\rm [S\,II] \lambda 6717+ \lambda 6731} /I_{{\rm H}\beta }$,  \\
$R_3$  = $I_{{\rm [O\,III]} \lambda 4959+ \lambda 5007} /I_{{\rm H}\beta }$.  \\
The Sun's mass is used as the unit for the masses of galaxies. 

\section{Data}

Thanks to deep, wide-field sky surveys the number of galaxies with
multi-wavelength data including spectra has increased considerably in
recent years.  In particular, numerous galaxies were measured by the
Sloan Digital Sky Survey, SDSS,
\citep{York2000,Stoughton02,Ahn12}.  Our investigation relies on the
SDSS data base (data release 12, DR12; \citep{Alam15}).  An important
task in our investigation is to select target galaxies with reliable
data (especially regarding masses and abundances) and to reject the
unreliable ones. 

\subsection{Our sample of ``environment galaxies'', $S_{EG}$}

\begin{figure}
\resizebox{1.00\hsize}{!}{\includegraphics[angle=000]{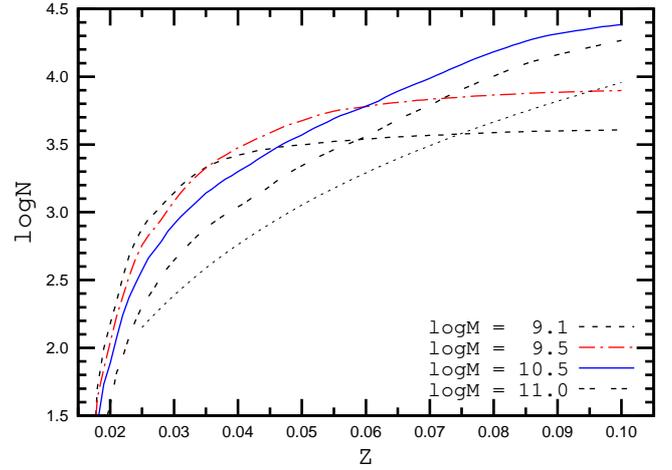}}
\caption{Cumulative number of galaxies with a redshift less than a
given value.  The curves for galaxies of four different values of the
stellar mass are shown.  The dotted curve shows the cumulative number
of galaxies as a function of redshift for the case of constant galaxy
space density. 
}
\label{figure:z-ngal-m}
\end{figure}

The  positions, redshifts, and stellar masses of 2,235,900 galaxies
were extracted from the SDSS data base (DR12, \citep{Alam15}).  We
adopted the spectral determination of the stellar masses of these
galaxies provided by the SDSS. This sample of galaxies is used for the
estimation of the environmental density and will be referred to as the
sample of ``environment galaxies'', $S_{EG}$. 

Fig.~\ref{figure:z-ngal-m} shows the cumulative number of galaxies
with a redshift less than a given value.  The curves for galaxies of
four different values of their stellar mass are shown. For each fixed
value of log~$M_{j}$ the number of galaxies within a mass interval from
log~$M_{j}$-0.05 to log~$M_{j}$+0.05 is determined.  The dotted curve
shows the cumulative number of galaxies as a function of redshift for
the case of constant galaxy space density. The shape of this curve is
used as reference. The curve for galaxies of a given mass  follows to
this shape until it reaches the redshift where the sample of galaxies
of this mass becomes incomplete, i.e., the comparison between the
shape of this curve and the shape of the curve for galaxies of a given
mass allows one to estimate the limiting redshift out to which the
galaxy sample of given mass interval is complete.  Inspection of
Fig.~\ref{figure:z-ngal-m} suggests that a realistic density of
galaxies with a mass of $10^{9.1}$ M$_{\sun}$ can be estimated out to
a redshift of $\sim$0.035 only, while for galaxies of mass of
$10^{10.5}$ M$_{\sun}$ it can be estimated out to $z \sim 0.1$.

Distances to the galaxies are calculated from 
\begin{equation}
d = \frac{c}{H_0} \times z,
\label{equation:r}   
\end{equation}
where $d$ is the distance in Mpc,  $c$ the speed of light in km
s$^{-1}$, and $z$ the redshift.  The value of the Hubble constant
$H_0$ was adopted to be 72 km s$^{-1}$ Mpc$^{-1}$. This value
was derived by the {\it Hubble Space Telescope} Key Project to measure
the Hubble constant \citep{Freedman2001}.  The analysis of the Planck
observations of temperature and polarization anisotropies of the
cosmic microwave backgraund results in a similar value of the  Hubble
constant, between 67 and 68 km s$^{-1}$ Mpc$^{-1}$
\citep{Ade2014,Ade2015}, as does the recent combination of Cepheid and
supernova Type Ia measurements for the local Hubble constant, yielding
$\sim 73$ km s$^{-1}$ Mpc$^{-1}$ \citep{Riess16}.

\subsection{Our sample of ``target galaxies'', $S_{TG}$}

Unfortunately, reliable abundances cannot be determined for all the SDSS galaxies.  
The gas-phase oxygen and nitrogen abundances of a star-forming galaxy can be 
estimated from emission line spectra. 
Spectra of $\sim 620,000$ galaxies with measurements of the necessary emission
lines H$\beta$, H$\alpha$, [O\,{\sc ii}]$\lambda \lambda$3727,3729,
[O\,{\sc iii}]$\lambda$5007, [N\,{\sc ii}]$\lambda$6584, [S\,{\sc
ii}]$\lambda$6717 and [S\,{\sc ii}]$\lambda$6731 are available in 
SDSS DR12. We select a sample of galaxies with reliable estimates 
of chemical abundances and will examine the influence of the environment 
on their abundances.  This sample of galaxies will be referred to as the 
sample of ``target galaxies'', $S_{TG}$. 
The sample of target galaxies is a subsample of environment galaxies, 
i.e. only a fraction of the SDSS galaxies (with reliable estimations of 
the abundances) is used as target galaxies while  all the SDSS galaxies 
are used to determine the environment density.

We used the hydrogen lines H$\beta$ and H$\alpha$ for the
de-reddening.  Specifically, we corrected the emission-line fluxes
for interstellar reddening using the theoretical
H$\alpha$/H$\beta$ ratio and the \citet{Whitford1958}  interstellar
reddening law (adopting the approximation suggested by
\citet{Izotovetal1994}). In several cases, the derived value of the
extinction $C$(H$\beta$) is negative and has then been set to zero.

The other lines were used for the determination of the oxygen
and nitrogen abundances through our calibration relations. 
Only the spectra where S/N $>$ 3 for the relevant emission lines 
are considered. It should be noted, however, that we also used additional criterion 
to select galaxies with reliable oxygen and nitrogen abundances (see below).
We then utilise the resulting abundances of this galaxy sample to examine the
influence of the environment, i.e., galaxy density, on the galactic
metallicities. 

Galaxies with AGN-like spectra were excluded using the  [N\,{\sc
ii}]$\lambda$6584/H$\alpha$ vs.\ [O\,{\sc iii}]$\lambda$5007/H$\beta$
diagnostic diagram of \citet{Baldwin1981}.  We adopt the dividing line
between H\,{\sc ii} region-like  and AGN-like spectra suggested by
\citet{Kauffmann2003}. 

The resulting sample of galaxies with reliable estimates of chemical
abundances and stellar masses is needed in order to examine the
influence of the environment density on the galactic metallicities. In
the following, we analyse our sample of the target galaxies after
excluding the galaxies with unreliable estimates of stellar mass and
abundance.

\begin{figure}
\resizebox{1.00\hsize}{!}{\includegraphics[angle=000]{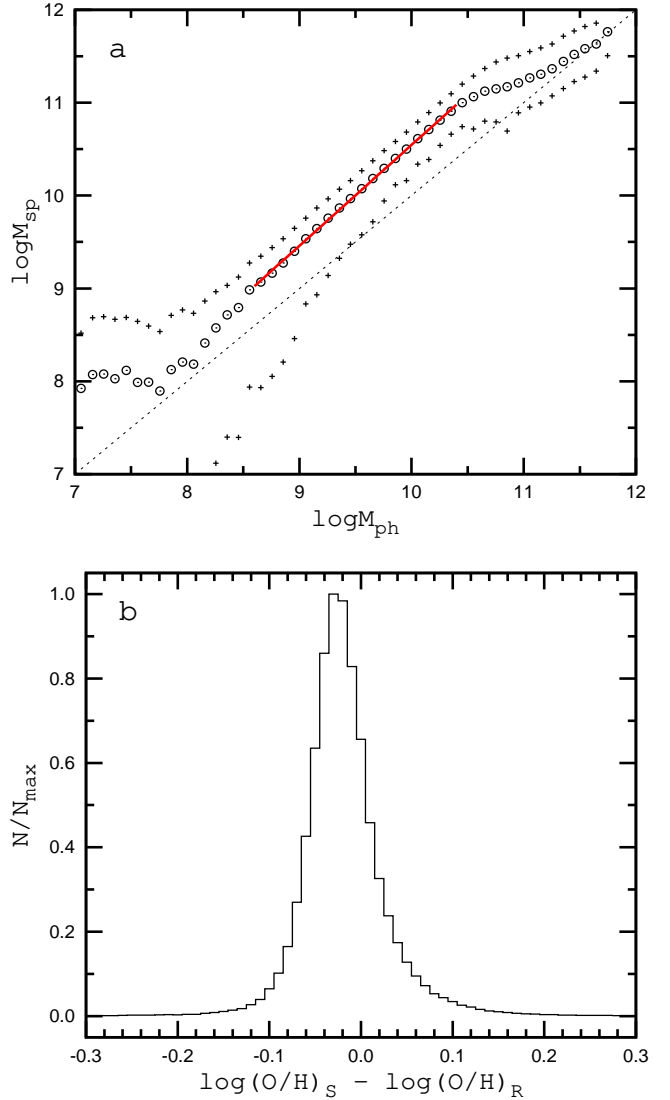}}
\caption{
Upper panel: The value of the spectroscopically derived mass of a
galaxy $M_{sp}$ as a function of the photometrically inferred mass
$M_{ph}$. The open circles are the mean values of log~$M_{ph}$ and
log~$M_{sp}$ in bins of 0.1 in log~$M_{ph}$, the plus signs are the
mean deviations from the mean values.  The solid line is the adopted
relation between  log~$M_{ph}$ and log~$M_{sp}$ within the mass
interval from  log~$M_{sp} =9$ to log~$M_{sp} =11$. The diagonal
dotted line is the equal value line.  The lower panel shows a
normalised histogram of the differences between the
$R$-calibration-based and $S$-calibration-based oxygen abundances
log(O/H)$_{S} -$~log(O/H)$_{R}$ in our galaxies. 
}
\label{figure:otbor}
\end{figure}

\subsubsection{The stellar masses of our galaxies}

The SDSS data base offers values of the stellar masses of galaxies
determined in different ways.  We have chosen the photometrical
$M_{ph}$ and spectral $M_{sp}$ masses of the SDSS and BOSS galaxies
(BOSS stands for the Baryon Oscillation Spectroscopic Survey in
SDSS-III, see \citet{Dawson13}).  The photometric masses are the
best-fit stellar masses from database table {\sc
stellarMassStarformingPort} obtained by the Portsmouth method, which
fits stellar evolution models to the SDSS photometry
\citep{Maraston2009}.  The spectral masses are the median (50th
percentile of the probability distribution function, PDF) of the
logarithmic stellar masses from table {\sc stellarMassPCAWiscBC03}
determined by the Wisconsin method \citep{Chen2012} with the stellar
population synthesis models from \citet{Bruzual2003}. 

The upper panel of Fig.~\ref{figure:otbor} shows the photometric mass
$M_{ph}$ as a function of the corresponding spectral mass $M_{sp}$.
The range of galaxy masses is devided into bins of 0.1 dex in
$M_{ph}$.  For the galaxies within each bin, the mean values of
log~$M_{ph}$ and  log~$M_{sp}$ and the mean deviations from the mean
values are determined.  The mean values of log~$M_{ph}$ and
log~$M_{sp}$  are shown in the upper panel of Fig.~\ref{figure:otbor}
by open circles, and the mean deviations by plus signs.
 
The upper panel of Fig.~\ref{figure:otbor} shows that the values of
log~$M_{ph}$ and log~$M_{sp}$ are roughly the same only for very
massive galaxies and that there is a shift between log~$M_{ph}$ and
log~$M_{sp}$ for galaxies of lower masses.  It should be noted that
\citet{Kannappan2007} have demonstrated that different stellar mass
estimation methods yield relative mass scales that can disagree by
factor $\ga 3$.  The relation between log~$M_{ph}$ and log~$M_{sp}$
can be approximated by the linear relation 
\begin{equation}
\log M^{*}_{sp} = 1.086 \log M_{ph} - 0.313 
\label{equation:msp-mph}
\end{equation}
for galaxies within the stellar mass interval $9 <$ log~$M_{sp} < 11$.
This relation can then be used to exclude galaxies with unreliable
stellar masses within this interval. More precisely, we exclude the
galaxies where the mean difference between  log~$M^{*}_{sp}$ and
log~$M_{sp}$ is larger than 0.2 dex. 

If the difference between  log~$M^{*}_{sp}$ and log~$M_{sp}$ can be interpreted 
as the random error of the stellar galaxy mass and the difference between 
log~$M_{ph}$ and log~$M_{sp}$ as a systematic error then the mean random error 
of the stellar galaxy mass for the sample of galaxies  
within the stellar mass interval $9 <$ log~$M_{sp} < 11$  is 
$\sim$0.11 dex and the systematic error can excess 0.5 dex.

In the following we will use the spectral masses of the galaxies and
will denote them as $M$ without an index.

\subsubsection{The chemical abundances of our galaxies}

The oxygen and nitrogen abundances of the galaxies of our sample are
estimated using two calibrations from \citet{Pilyugin2016} where two
sets of strong emission lines are used. The oxygen abundances
(O/H)$_{R}$ are determined using the $R_2$, $R_3$, and $N_2$ line
intensities  
\begin{eqnarray}
       \begin{array}{lll}
     {\rm (O/H)}^{*}_{R}  & = &   8.589 + 0.022 \, \log (R_{3}/R_{2}) + 0.399 \, \log N_{2}   \\  
                          & + &  (-0.137 + 0.164 \, \log (R_{3}/R_{2}) + 0.589 \log N_{2})   \\ 
                          & \times &  \log R_{2}   \\ 
     \end{array}
\label{equation:ohru}
\end{eqnarray}
if  log~$N_{2} > -0.6$, and  
\begin{eqnarray}
       \begin{array}{lll}
     {\rm (O/H)}^{*}_{R}  & = &   7.932 + 0.944 \, \log (R_{3}/R_{2}) + 0.695 \, \log N_{2}   \\  
                          & + &  (0.970 - 0.291 \, \log (R_{3}/R_{2}) - 0.019 \log N_{2})   \\ 
                          & \times & \log R_{2}   \\ 
     \end{array}
\label{equation:ohrl}
\end{eqnarray}
if log~$N_{2} < -0.6$. The notation (O/H)$^{*}$ = 12 +log(O/H) is used
for the sake of the brevity. 

The oxygen abundances (O/H)$_{S}$ are determined using the $N_2$,
$R_3$, and $S_2$ line intensities 
\begin{eqnarray}
       \begin{array}{lll}
     {\rm (O/H)}^{*}_{S}  & = &   8.424 + 0.030 \, \log (R_{3}/S_{2}) + 0.751 \, \log N_{2}   \\  
                          & + &  (-0.349 + 0.182 \, \log (R_{3}/S_{2}) + 0.508 \log N_{2})   \\ 
                          & \times & \log S_{2}   \\ 
     \end{array}
\label{equation:ohsu}
\end{eqnarray}
if log~$N_{2} > -0.6$, and  
\begin{eqnarray}
       \begin{array}{lll}
     {\rm (O/H)}^{*}_{S}  & = &   8.072 + 0.789 \, \log (R_{3}/S_{2}) + 0.726 \, \log N_{2}   \\  
                          & + &  (1.069 - 0.170 \, \log (R_{3}/S_{2}) + 0.022 \log N_{2})    \\ 
                          & \times & \log S_{2}   \\ 
     \end{array}
\label{equation:ohsl}
\end{eqnarray}
if log~$N_{2} < -0.6$. 

A normalised histogram of the differences between those values of the
abundance log(O/H)$_{S} -$ log(O/H)$_{R}$ is presented in the lower
panel of Fig.~\ref{figure:otbor}.  The scatter in the differences of
oxygen abundances is rather small but there is a systematic shift of
around $-0.03$ dex in the values of log(O/H)$_{S}$ in comparison to
the values of log(O/H)$_{R}$. We determine the linear relation between
log(O/H)$_{S}$ and log(O/H)$_{R}$ 
\begin{equation}
12 + \log({\rm O/H})^{c}_{R} = 12 + \log({\rm O/H})_{S} + 0.029
\label{equation:ohr-ohs}
\end{equation}
and use this relation to exclude galaxies with unreliable estimations
of oxygen abundance. The adopted selection criterion is the following.
Galaxies where the absolute value of the
difference between log(O/H)$^{c}_{R}$ and log(O/H)$_{R}$ exceeds 0.05
are excluded from the sample.  This means that we exclude a galaxy if
its emission line [O\,{\sc ii}]$\lambda \lambda$3727,3729 or/and its
emission line [S\,{\sc ii}]$\lambda \lambda$6717,6731 measurements
show large uncertainties. It should be emphasised that it is difficult
to reveal the uncertainties in the measurements of the emission line
[O\,{\sc iii}]$\lambda \lambda$4959,5007 and [N\,{\sc ii}]$\lambda
\lambda$6548,6584 through comparison of the abundances  log(O/H)$_{S}$
and log(O/H)$_{R}$ since those lines can contribute similar errors in
both. Strictly speaking, we excluded only a fraction of galaxies with
unreliable oxygen abundances. 

Below only the oxygen abundances (O/H)$_{R}$ will be used, and we will
refer to them as O/H. 
 
The nitrogen abundances were estimated in two ways.  First, the
nitrogen abundances were obtained through the following relations
\citep{Pilyugin2016}:
\begin{eqnarray}
       \begin{array}{lll}
     {\rm (N/H)}^{*}_{R}  & = &   7.939 + 0.135 \, \log (R_{3}/R_{2}) + 1.217 \, \log N_{2}   \\  
                          & + &  (-0.765 + 0.166 \, \log (R_{3}/R_{2}) + 0.449 \log N_{2})   \\ 
                          & \times & \log R_{2}   \\ 
     \end{array}
\label{equation:nhru}
\end{eqnarray}
if log~$N_{2} > -0.6$, and
\begin{eqnarray}
       \begin{array}{lll}
     {\rm (N/H)}^{*}_{R}  & = &   7.476 + 0.879 \, \log (R_{3}/R_{2}) + 1.451 \, \log N_{2}   \\  
                          & + &  (-0.011 - 0.327 \, \log (R_{3}/R_{2}) - 0.064 \log N_{2})   \\ 
                          & \times & \log R_{2}   \\ 
     \end{array}
\label{equation:nhrl}
\end{eqnarray}
if log~$N_{2} < -0.6$. Again the notation (N/H)$^{*} = 12 +$~log(N/H)
is adopted for the sake of brevity. 

Second, the nitrogen-to-oxygen abundance ratio N/O is estimated using
the expression that relates the nitrogen-to-oxygen abundance ratio N/O
in an H\,{\sc ii} region with the intensities of the strong lines in
its spectrum, 
\begin{eqnarray}
       \begin{array}{lll}
\log {\rm (N/O)}  & = &  -0.657 - 0.201 \, \log N_{2}  \\  
                  & + & (0.742 -0.075 \, \log N_{2}) \times \log(N_{2}/R_{2})  \\ 
     \end{array}
\label{equation:nolin}
\end{eqnarray}
and then the nitrogen abundance is determined using 
\begin{equation}
\log {\rm (N/H)} =  \log {\rm (O/H)} + \log {\rm (N/O)} 
\label{equation:nh2}
\end{equation}
where the oxygen abundance (O/H)$_{R}$ is used.  The galaxies where
the absolute value of the difference between two values of nitrogen
abundances exceeds 0.05 dex are excluded from the sample 
(one more selection criterion).

Below, only nitrogen abundances estimated in the second way will be
used, and we will refer to them as N/H.

\subsubsection{Metallicity -- stellar mass diagram}

\begin{figure}
\resizebox{1.00\hsize}{!}{\includegraphics[angle=000]{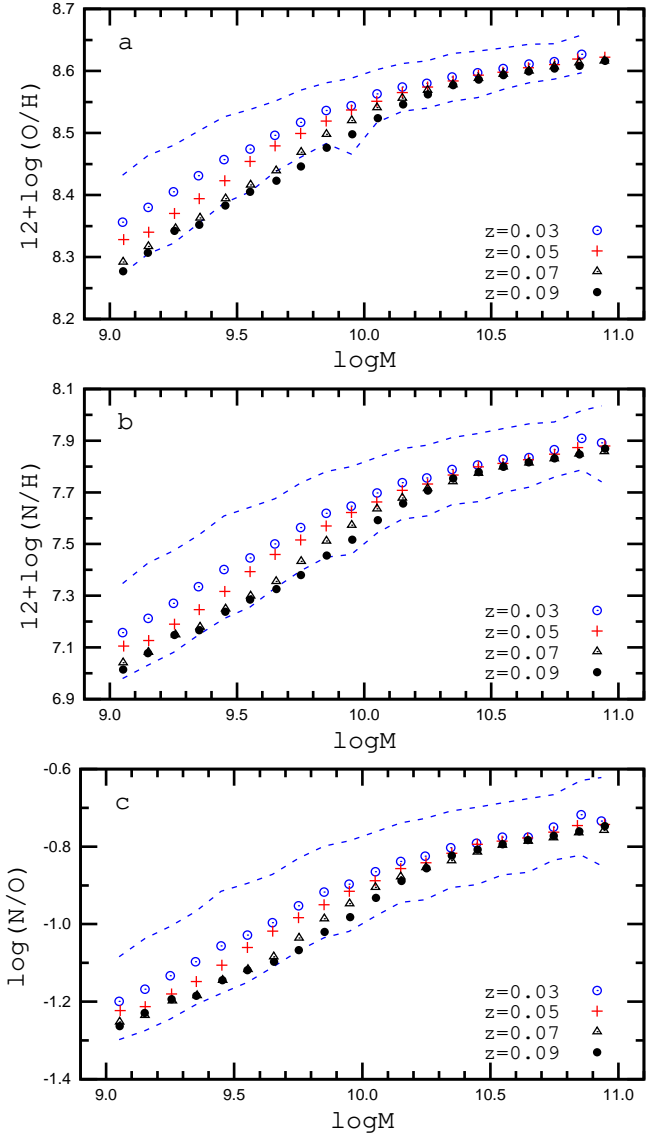}}
\caption{
Panel $a.$ The oxygen abundance vs.\ stellar mass diagram for our
sample of target galaxies.  The O/H -- $M$ relations for galaxies in
four redshift intervals of $z \pm 0.01$ are presented.  The symbols
indicate the mean values for the bins of 0.1 dex in stellar masses.
The dashed lines show the mean deviations for the subsample of 
galaxies with redshift $z$ = 0.03. 
Panels $b$ and $c$ show the same as panel $a$ but for nitrogen 
and nitrogen-to-oxygen ratio, respectively.
}
\label{figure:m-xh-z}
\end{figure}

Our sample of target galaxies with stellar masses in the range of $9.1
<$ log$M < 11$ and redshifts  $z < 0.1$ after excluding AGNs and
galaxies with unreliable estimates of stellar masses and oxygen and
nitrogen abundances amounts to 77,659 galaxies.  

Panel $a$ of Fig.~\ref{figure:m-xh-z} shows the O/H -- $M$ diagram for this sample
of target galaxies.  The O/H -- $M$ relations for galaxies in four
redshift bins $z_{j} \pm 0.01$ (where $z_{j} = 0.03$, 0.05, 0.07, and
0.09) are presented.  The symbols in Fig.~\ref{figure:m-xh-z}  show
the mean values in bins of 0.1 dex of the stellar masses.  The O/H --
$M$ relations at low redshifts are more reliable for the low-mass
galaxies than for the massive galaxies because the low-mass galaxies
are numerous at low redshifts in our sample while the number of
massive galaxies is rather low,  see Fig.~\ref{figure:z-ngal-m}.
Conversely, the O/H -- $M$ relations for large (near $z \sim 0.1$)
redshifts are more reliable for massive galaxies than for the low-mass
galaxies.

The dashed lines in panel $a$ of Fig.~\ref{figure:m-xh-z} show the mean 
deviations of the oxygen abundances for the subsample of galaxies with redshift $z$ = 0.03. Inspection of 
panel $a$ of Fig.~\ref{figure:m-xh-z} shows that the change in the oxygen abundances 
in low-mass galaxies due to the evolution over the redshift diapason $\sim$0.05  
(over the time interval $\sim$0.5 Gyr) is comparable to the scatter in abundances 
for galaxies at the fixed redshift $z$ = 0.03. Therefore, a deviation of the abundance of an individual 
galaxy from the mass -- metallicity relation can serve as indicator of overabundance 
(in particular, due to influence of the environment) if only the galaxies within a small 
diapason of redshift are considered, i.e. if the evolutionary effect on the galaxy 
oxygen abundance is excluded. 
The evolutionary change of the oxygen abundance in massive galaxies at the 
considered redshift interval is less pronounced.

Panels $b$ and $c$ show the same as panel $a$ but for the nitrogen abundance 
and the nitrogen-to-oxygen abundances ratio, respectively.
Again, the change in the nitrogen abundances and nitrogen-to-oxygen 
abundances ratio in low-mass galaxies due to the evolution 
over the time interval $\sim$0.5 Gyr is comparable to the scatters 
for galaxies at the redshift $z$ = 0.03.

The examination of the X/H -- $M$ diagrams in
Fig.~\ref{figure:m-xh-z} reveals three major trends of galactic
chemical abundances with mass and redshift.  \\
{\it (i)}. The oxygen abundance increases with increasing masses and
with decreasing redshift.  The X/H -- $M$ relation flattens at the
massive end.  This agrees with the mass-metallicity (or
luminosity-metallicity) relation of galaxies revealed by
\citet{Lequeux1979} and confirmed in the local universe and at high
redshifts in many earlier studies \citep[e.g.,][]{Zaritsky1994,
Garnett2002, Grebel2003, Tremonti2004, Erb2006, Pilyugin2007,
Maiolino2008, Guseva2009, Pilyugin2013, Maier2014, Izotov2015}.     \\
{\it (ii)}. The rate of the change of oxygen abundances with redshift
decreases with increasing galaxy masses.  \citet{Gavazzi1993} found
that the present-day rate of evolution of late-type galaxies decreases
with increasing mass, suggesting that the efficiency of conversion of
gas into stars at an early epoch increases with the mass of the
system.  This effect has been named ``downsizing'' and has been
confirmed in many studies \citep[e.g.,][]{Cowie1996, Heavens2004,
Maiolino2008, Tomczak2016}.  \\
{\it (iii)}.  The rate of the change of nitrogen abundance with
redshift and stellar mass is higher than that for oxygen. The faster
rate of the change of nitrogen abundances as compared to oxygen
abundances above $12 +$ log(O/H) $\sim 8.2$ clearly demonstrates the
rise of N/O in the  N/O vs.\ O/H diagram considered in many studies
\citep[][among many
others]{Edmunds1978,Izotov1999,Henry2000,Pilyugin2003,Pilyugin2004,Berg2012,Annibali2015,Croxall2016}. \\

Since only a sample of galaxies within a rather small interval of redshifts (ages) 
are considered, the evolutionary changes in oxygen abundances are small, 
around 0.05 dex. Nevertheless, our X/H~$=f(z,M)$ relations reproduce the known 
trends in galactic abundances.
This can be considered as evidence that the precision of 
the derived abundances is high enough to allow us to detect 
abundance changes of the order of $\sim$0.05 dex and below.

\begin{figure}
\resizebox{1.00\hsize}{!}{\includegraphics[angle=000]{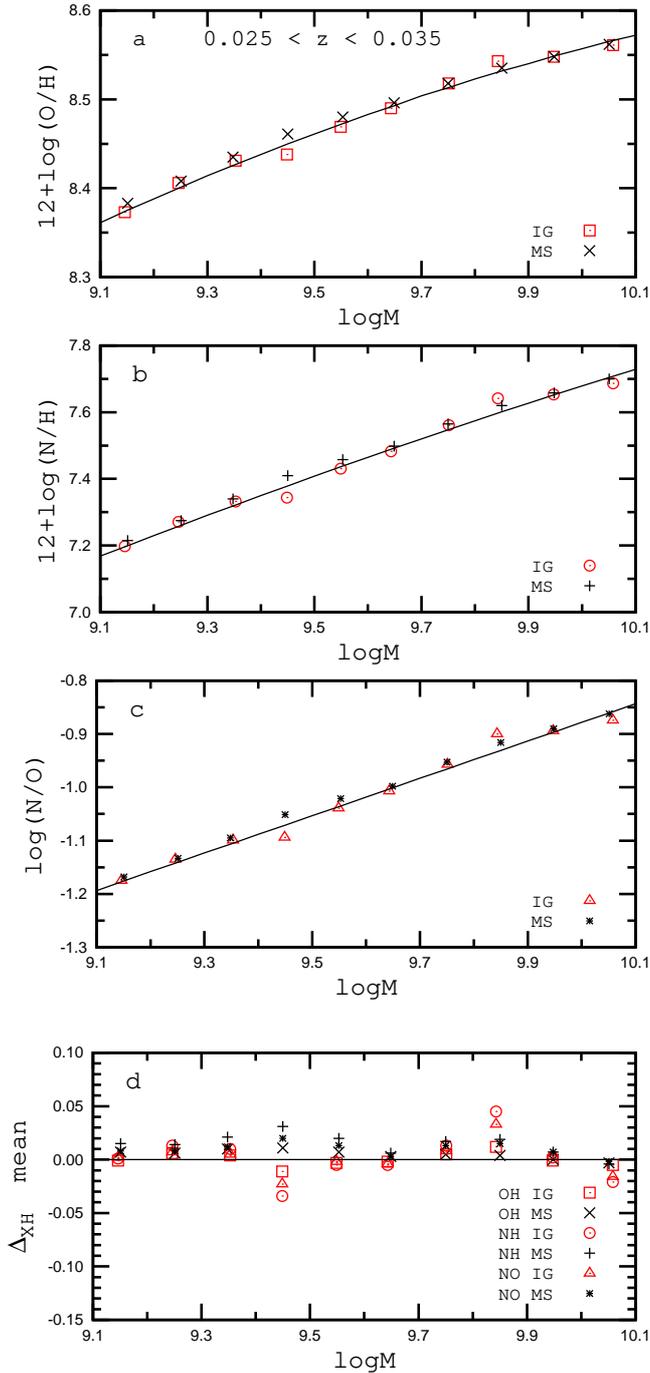}}
\caption{Comparison between the abundances of isolated galaxies (IGs)
and galaxies with neighbours (multiple-system galaxies, MSs).  
Panel $a.$ The solid line represents the O/H -- $M$ relation for IGs
(Eq.~\ref{equation:oh-z030}).   The squares show the mean oxygen
abundances for the IGs in bins of 0.1 dex in stellar masses. The
crosses indicate the mean oxygen abundances for MSs.  
Panel $b.$  The solid line is the N/H -- $M$ relation for IGs
(Eq.~\ref{equation:nh-z030}).    The circles denote the mean nitrogen
abundances for the IGs in bins of 0.1 dex in stellar masses. The plus
signs mark the mean nitrogen abundances for MSs.  
Panel $c.$  The solid line is the N/O -- $M$ relation for IGs
(Eq.~\ref{equation:no-z030}).    The triangles are the mean 
nitrogen-to-oxygen abundance ratios for the IGs in bins of 0.1 dex in 
stellar masses. The asterisks show the mean nitrogen abundances for MSs.  
Panel $d.$ The mean
chemical overabundance of galaxies in bins of 0.1 dex in stellar
masses.  The overabundances in oxygen are shown by squares for the IGs
and by crosses for the MSs.  The overabundances in nitrogen are shown
by circles for the IGs and by plus signs for the MSs. 
The deviations in nitrogen-to-oxygen ratio are shown
by triangles for the IGs and by asterisks for the MSs. 
}
\label{figure:ig-ms-z030}
\end{figure}

The SDSS spectra are measured through 3 arcsec diameter fibres and, 
consequently, at low redshifts the projected aperture diameter is smaller than 
that at high redshift.  
Since there is usually a radial abundance gradient in galaxy discs
\citep[see][]{VilaCostas1992,Zaritsky1994,vanZee1998,
Pilyugin2004,Pilyugin2006,Pilyugin2014,Moustakas2010,Sanchez2014},
an aperture-redshift effect can be present in
abundances based on SDSS spectra \citep[e.g.][]{Kewley2005}. 
At low redshift the SDSS spectra are the spectra of the central parts of 
galaxies while at large redshift they are close to global spectra of 
whole galaxies, i.e. the abundances based on the SDSS spectra
are more biased towards the central metallicity at low than at high redshifts. 
The aperture effect can contribute to the 
mass-dependent redshift evolution seen in Fig.~\ref{figure:m-xh-z}. 
To exclude (or at least minimise) the aperture effect on the scatter 
in abundances, a sample of galaxies within a small interval of redshifts 
should be considered.

\section{Influence of the environment on the chemical abundances in 
galaxies}

\subsection{Environmental influence on the abundances in 
galaxies of masses of $10^{9.1}$~M$_{\sun}$ to $10^{10.1}$~M$_{\sun}$}

We noted above that a realistic environmental density of galaxies of
masses of log~$M = 9$ can be estimated out to a redshift of
$\sim0.035$ only (see Fig.~\ref{figure:z-ngal-m}).  Here, we examine
now the influence of the environment on the chemical abundances of
target galaxies with redshifts from $z = 0.025$ to 0.035. Since the
redshift interval is small the evolutionary change (i.e., change with
redshift) of the abundance can be neglected.  We will also restrict
ourselves to the examination of target galaxies within the stellar
mass range from log~$M = 9.1$ to 10.1. On the one hand, those galaxies
are the most numerous in our sample of the target galaxies at redshifts
from $z = 0.025$ to 0.035. On the other hand,  one can expect (see
Fig.~\ref{figure:m-xh-z}) that the X/H -- $M$ relations for the
galaxies within this stellar mass interval can be well reproduced by a
simple expression.  This sample amounts to 4572 galaxies and will be
referred to as sample $S030$.

\subsubsection{Isolated galaxies vs.\ multiple systems}

We divide sample $S030$ into two subsamples: the subsample of isolated
galaxies (IG) and the subsample of multiple systems (MS), i.e.,
galaxies with neighbours.  To select the isolated galaxies, a criterion
based on a selection in a projected distance -- velocity difference
space is usually employed. However, different numerical values for
such criteria are adopted in the literature
\citep[e.g.,][]{Fuse2012,Argudo2015,Spector2016,Lacerna2016}.  We have
chosen to use the isolation criteria from \citet{Lacerna2016}. A
galaxy is considered as an IG if it has no known neighbour within the
projected separation across the line of sight $R$ of less than 100
typical isophotal radii $R_{25}$ and with a radial velocity difference
$\Delta$$V$ less than 1000 km s$^{-1}$. We adopt the typical value of
$R_{25}$  for late-type galaxies to be 13 kpc \citep{Pilyugin2014}.
Then the isolation criteria are 
\begin{eqnarray}
       \begin{array}{lll}
    R     & > & 1.3 \; \: {\rm Mpc}    \\  
          &   &             \\  
\Delta z  & > &  0.0033   \\ 
     \end{array}
\label{equation:ic}
\end{eqnarray}

For each target galaxy from sample $S030$, we search for neighbours in
our sample of environment galaxies $S_{EG}$.  It should be emphasised
that only environment galaxies with masses higher than $10^{9}$
M$_{\sun}$ were taken into account.  531 galaxies from the sample
$S030$ are isolated galaxies according to our isolation criteria. 

For isolated galaxies, we obtain the O/H -- $M$ relation 
\begin{equation}
12 + \log({\rm O/H}) = 8.333 + 0.290 \,m - 0.0658 \, m^{2} ,
\label{equation:oh-z030}
\end{equation}
the N/H -- $M$ relation 
\begin{equation}
12 + \log({\rm N/H}) = 7.104 + 0.640 \,m - 0.0660 \, m^{2} ,
\label{equation:nh-z030}
\end{equation}
and the N/O -- $M$ relation 
\begin{equation}
\log({\rm N/O}) = -1.228 + 0.351 \,m - 0.0003 \, m^{2} ,
\label{equation:no-z030}
\end{equation}
where $m =$ log~$M - 9$.  A few galaxies (14 out of 531) show large
deviations from the O/H -- $M$ ($d_{OH} > 0.15$ dex) or from the N/H
-- $M$ ($d_{NH} > 0.40$ dex) relations. Those galaxies are not used in
deriving the final relations and are excluded from our further
analysis.  The mean deviation from the final relations (i.e., the
mean value of the residuals in the relations) 
\begin{equation}
d_{XH} = \left(\frac{1}{n}\sum_{j=1}^{n} (\log ({\rm X/H})^{OBS}_{j} - \log ({\rm X/H})^{CAL}_{j})^{2} \right)^{1/2}  
\label{equation:delta}
\end{equation}
is $d_{OH} = 0.058$ dex, $d_{NH} = 0.160$ dex, and $d_{NO} = 0.106$ dex 
for the remaining 517 isolated galaxies. 

It is widely accepted that the mass -- metallicity relation for
galaxies is caused by the variation of the astration level (or gas
mass fraction) with galaxy mass, in the sense that the astration level
increases (gas mass fraction decreases) with increasing galaxy mass.
The scatter in abundances among the galaxies of a given stellar mass
evidences that there is a spread in the astration levels (or gas mass
fractions) among those galaxies. Surely, the uncertainties in the
stellar masses and the abunances may be responsible for the some
fraction of the scatter in abundances. The variation with  astration
level is larger for nitrogen abundances than for oxygen because the
secondary nitrogen is dominant at the metallicities in our sample of
galaxies. As a consequence, the scatter in abundances among galaxies
of a given mass due to the spread in the astration levels is larger
for nitrogen than for oxygen.  The time delay between nitrogen and
oxygen enrichment together with the different star formation histories
in galaxies can make an additional contribution to the scatter in the
nitrogen abundances
\citep[e.g.,][]{Edmunds1978,Pagel1997,Henry2000,PilyuginThuan2011}.

Let us compare the agreement of the X/H -- $M$ diagrams for the
isolated galaxies and for multiple system galaxies.  The O/H -- $M$
relation obtained for the isolated galaxies of masses in the range
$10.1 \ga$ log~$M \ga 9.1$ at redshifts $0.035 \ga z \ga 0.025$ is
shown by the solid line in panel $a$ of Fig.~\ref{figure:ig-ms-z030}.
The squares represent the mean oxygen abundances for IGs in bins of
0.1 dex in stellar mass. The crosses show the mean oxygen abundances
for MSs.  We refer to the difference between the measured abundance and 
the abundance obtained from the  X/H -- $M$ relation as
``overabundance''.  The mean chemical overabundance of a subsample of 
galaxies is defined by 
\begin{eqnarray}
       \begin{array}{lll}
\Delta_{XH}   & = & \frac{1}{n}\sum_{j=1}^{n} (\log ({\rm X/H})^{OBS}_{j} - \log ({\rm X/H})^{CAL}_{j})  \\
     \end{array}
\label{equation:exh}
\end{eqnarray}
where  X/H = O/H or X/H = N/H.  The mean oxygen overabundances
$\Delta_{OH}$ in bins of 0.1 dex in stellar mass are shown by squares
for the IGs and by crosses for the MSs in panel $d$ of
Fig.~\ref{figure:ig-ms-z030}.  The values of $\Delta_{OH}$ in the bins
depend on the choice of the bin size and are used for illustration
only.  The oxygen overabundance averaged over all MSs specifies the
mean shift of the abundances in MSs from the OH -- $M$ relation traced
by the isolated galaxies. The mean oxygen overabundance for all (3900)
MSs is $\Delta_{OH} = 0.0055$. 

The solid line in panel $b$ of Fig.~\ref{figure:ig-ms-z030} is the
obtained N/H -- $M$ relation. The circles mark the mean nitrogen
abundances for IGs in bins of 0.1 dex in stellar masses. The plus
signs denote the mean oxygen abundances for MSs.  The overabundances
in nitrogen are shown by circles for IGs and by plus signs for MSs in
panel $d$ of Fig.~\ref{figure:ig-ms-z030}. The mean nitrogen
overabundance for all MSs is $\Delta_{NH} = 0.0155$.  It should be
noted that \citet{Kacprzak2015} have also recently found that there is
no discernible difference between the mass -- metallicity relation of
field and cluster galaxies to within 0.02 dex.

The solid line in panel $c$ of Fig.~\ref{figure:ig-ms-z030} is the
obtained N/O -- $M$ relation. The triangles are the mean N/O 
ratios for IGs in bins of 0.1 dex in stellar masses. The asterisks 
denote the nitrogen-to-oxygen ratios for MSs.  The N/O deviations 
are shown by triangles for IGs and by asterisks for MSs in
panel $d$ of Fig.~\ref{figure:ig-ms-z030}. The mean N/O deviation 
for all MSs is $\Delta_{NH} = 0.0100$.

Thus, the general metallicity -- mass relations for isolated galaxies
and galaxies having neighbours are close to each other. The mean shift
of non-isolated galaxies around the metallicity -- mass relation
traced by the isolated galaxies is less than $\sim 0.01$ dex for
oxygen and less than $\sim 0.02$ dex for nitrogen.  This suggests that
environmental effects do not play a key role in the chemical evolution
of galaxies.

\begin{figure*}
\resizebox{1.00\hsize}{!}{\includegraphics[angle=000]{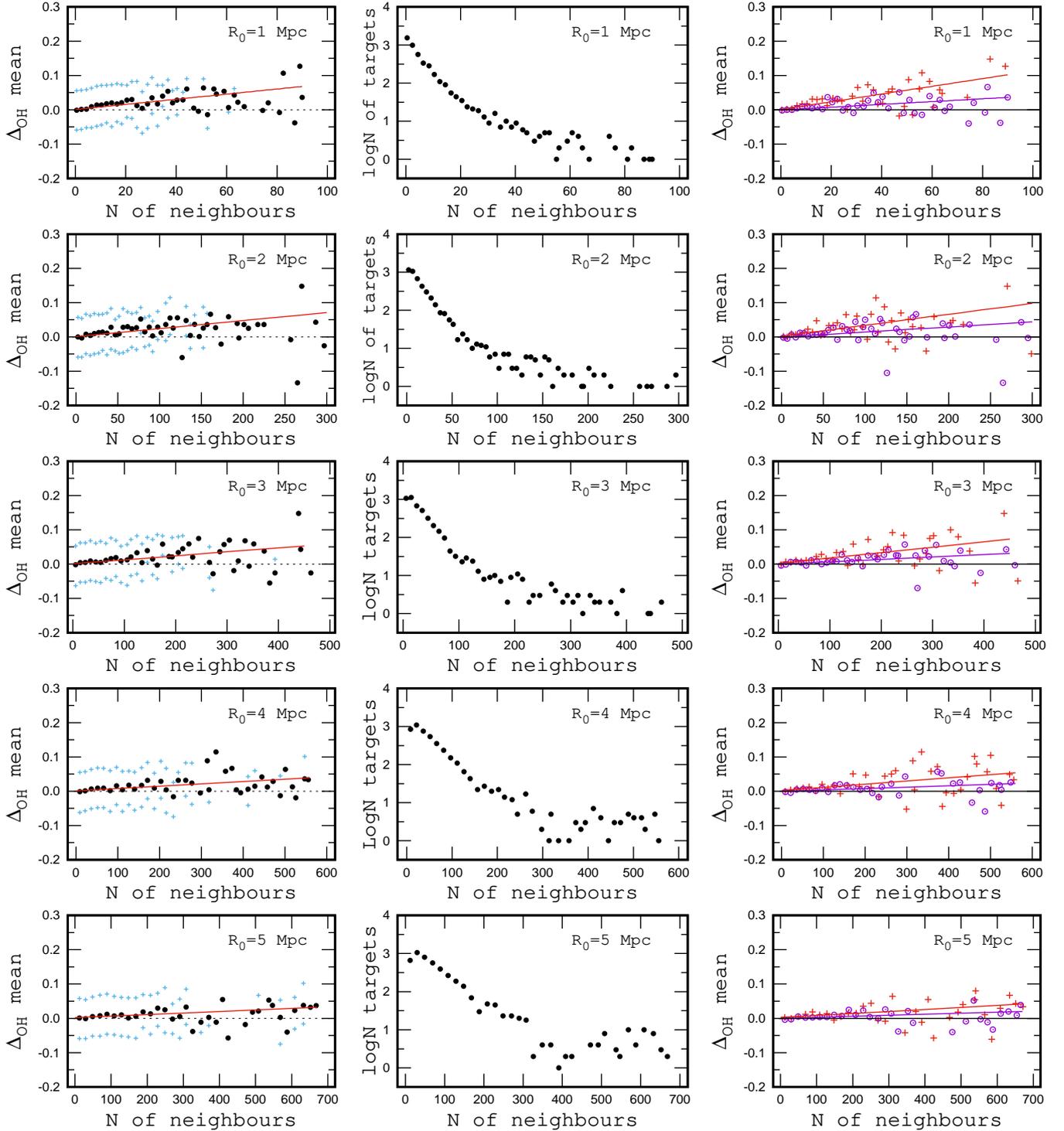}}
\caption{
Left column: Mean oxygen overabundance $\Delta_{\rm OH}$ (black
points) and mean deviation (blue plus signs) of the target galaxies
with stellar masses of $9.1 \la$ log$M \la 10.1$ at redshifts $0.025
\la z \la 0.035$ as a function of the number of neighbour galaxies for
different values of $R_{0}$ as labeled in each panel.  The solid line
is the linear best fit to all the individual data points.  The value
of the peculiar velocities of the neighbouring galaxies is adopted to be
equal to 1000 km s$^{-1}$.  The mean values are obtained for galaxies
in bins of the number of neighbouring galaxies. The bins are 2 for
$R_{0} = 1$ Mpc,  5 for $R_{0} = 2$ Mpc,  10 for $R_{0} = 3$ Mpc, 15
for $R_{0} = 4$ Mpc,  and 20 for $R_{0} = 5$ Mpc.  Middle
column:  The logarithm of the number of target galaxies having a
given number of neighbour galaxies.  Right column: The circles
indicate the mean oxygen overabundance of galaxies of $9.6 <$ log$M <
10.1$ and the solid line is the linear best fit.  The plus signs show
the mean oxygen overabundance of galaxies of masses $9.1 <$ log$M <
9.6$ and the dashed line is the linear best fit. 
}
\label{figure:n-doh-z03-v1000}
\end{figure*}

\begin{figure}
\resizebox{1.00\hsize}{!}{\includegraphics[angle=000]{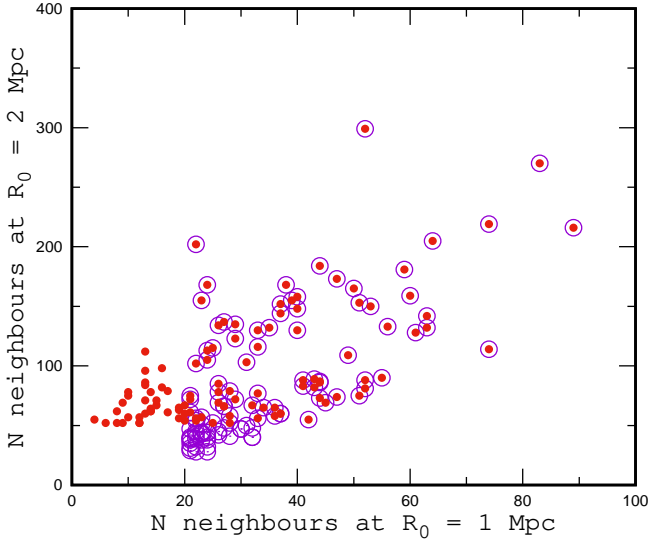}}
\caption{The relation between the number of neighbour galaxies at
$R_{0} = 2$ Mpc and at $R_{0} = 1$ Mpc for galaxies of $9.1 <$ log$M <
9.6$.  The circles visualise data for 111 galaxies with numbers of
neighbour galaxies at $R_{0} = 1$ Mpc higher than 20.  The points show
data for 113 galaxies with numbers of  neighbour galaxies at $R_{0} =
2$ Mpc higher than 51. 
}
\label{figure:n-n}
\end{figure}

\begin{figure*}
\resizebox{1.00\hsize}{!}{\includegraphics[angle=000]{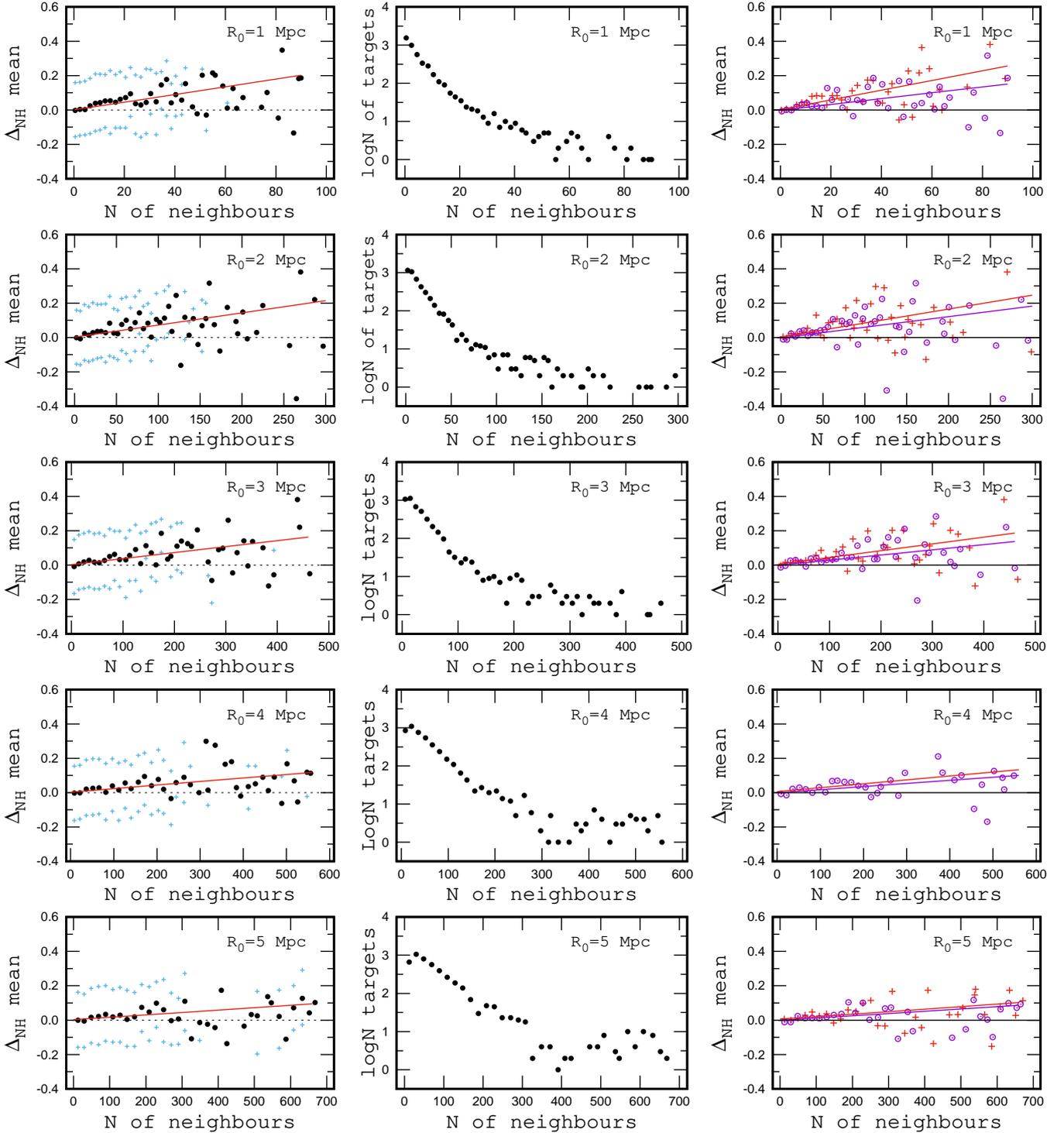}}
\caption{
The same as Fig.~\ref{figure:n-doh-z03-v1000} but for nitrogen. 
}
\label{figure:n-dnh-z03-v1000}
\end{figure*}

\begin{figure*}
\resizebox{1.00\hsize}{!}{\includegraphics[angle=000]{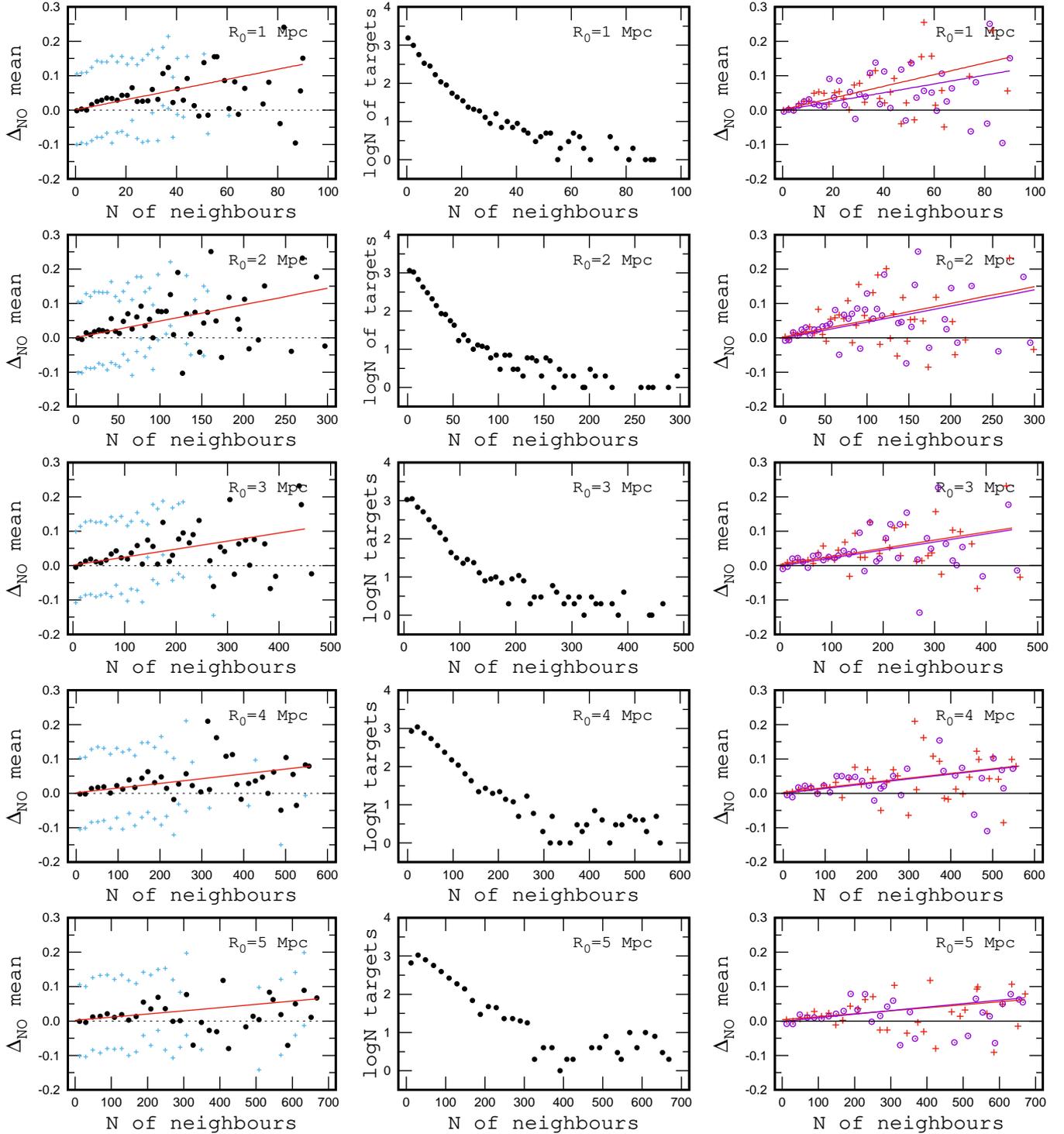}}
\caption{
The same as Fig.~\ref{figure:n-doh-z03-v1000} but for nitrogen-to-oxygen ratio.
}
\label{figure:n-dno-z03-v1000}
\end{figure*}

\begin{figure*}
\resizebox{1.00\hsize}{!}{\includegraphics[angle=000]{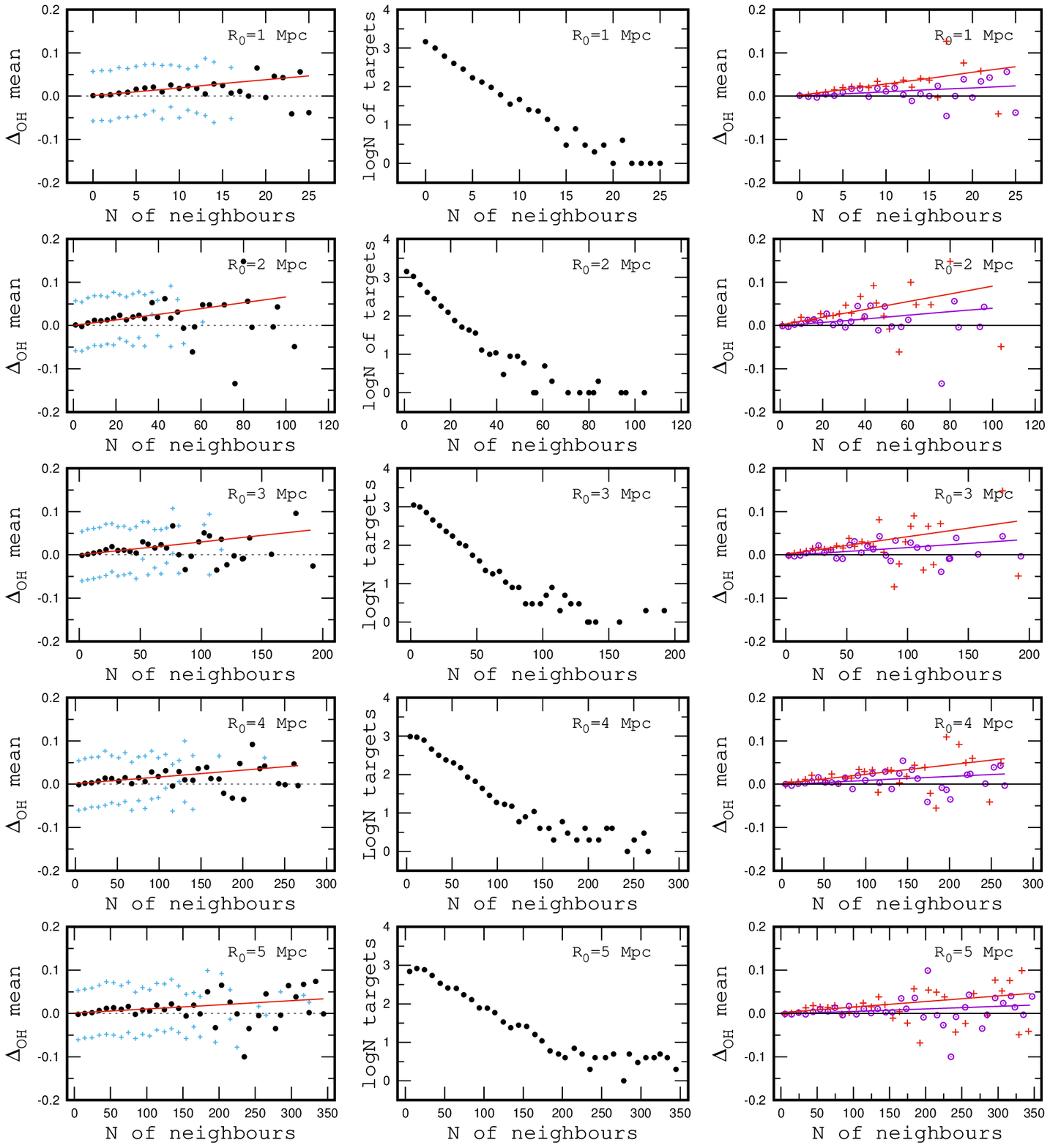}}
\caption{
The same as Fig.~\ref{figure:n-doh-z03-v1000} but for the value of 
the peculiar velocities of the neighbour galaxies of 100 km s$^{-1}$.
}
\label{figure:n-doh-z03-v0100}
\end{figure*}

\subsubsection{Chemical overabundance in galaxies as a function of 
number of neighbours}

Here we examine whether chemical overabundances in a galaxy depend on the
number of neighbour galaxies, i.e., on the environmental density. 

The environmental density of a galaxy can be quantified in different
ways. Here it will be specified by the number of galaxies within a
region of a certain projected distance -- velocity difference space.
The number of galaxies in the region is determined using neighbourhood
criteria defined in a way similar to the isolation criteria.  A galaxy
is considered to be a neighbour of the target galaxy if it has the
projected separation across the line of sight of less than a fixed
value $R_{0}$ and a redshift difference less than a fixed value
$dz_{0}$.  Five different values of $R_{0}$ are considered: $R_{0}  =
1$ Mpc, 2 Mpc, 3 Mpc, 4 Mpc, and 5 Mpc. Each value of $R_{0}$ is
accompanied by a value of $dz_{0}$ that involves two components. The
first component corresponds to the change of redshift with distance
equal to the adopted $R_{0}$.  The second component takes the peculiar
velocities of galaxies, $V_{p}$, into account.  Thus, the total value
of $dz_{0}$ is the sum $dz_{0} = dz_{0}(R_{0}) + dz_{0}(V_{p})$. Three
values of peculiar velocities $V_{p}$ are considered:  $V_{p} = 1000$
km s$^{-1}$, 500 km s$^{-1}$, 100 km s$^{-1}$.

Let us first discuss the case of $V_{p} = 1000$ km s$^{-1}$. For each
target galaxy from sample $S030$, we count the number of neighbour
galaxies in the sample of the environment galaxies $S_{EG}$.  The
panels in the left column of  Fig.~\ref{figure:n-doh-z03-v1000} show
the mean oxygen overabundances as a function of the number of neighbour
galaxies within the regions defined by $R_{0}$ specified in each
panel.  The target galaxies are divided into bins according to the
number of neighbour galaxies. The bin sizes are 2 for $R_{0} = 1$ Mpc,
5 for $R_{0} = 2$ Mpc,  10 for $R_{0} = 3$ Mpc,  15 for $R_{0} = 4$
Mpc,  and 20 for $R_{0} = 5$ Mpc.  The mean oxygen overabundances
$\Delta_{\rm OH}$ are determined for the target galaxies in each bin
and shown in the panels in the left column of
Fig.~\ref{figure:n-doh-z03-v1000} by the filled circles.  The solid
line is the linear best fit to all the individual galaxies (not to the
bin means shown in the figure).  The scatter in the overabundances
among galaxies in the bin is specified by the mean deviation of the
overabundances  $d$($\Delta_{\rm OH}$). The separate deviations above
and below the mean value of the overabundance are calculated if the
number of galaxies with deviations above (below) is at least 3. The
mean deviations of the overabundances are shown by plus signs in the
panels of the left column of Fig.~\ref{figure:n-doh-z03-v1000}.  

The panels of the middle column of Fig.~\ref{figure:n-doh-z03-v1000}
show the logarithm of the number of target galaxies having a given
number of neighbour galaxies. Inspection of these panels suggests that
there is some influence of the environment on the abundances, i.e.,
the galaxies in the dense environments tend to have an overabundance
of oxygen. 

Fig.~\ref{figure:n-doh-z03-v1000} suggests that the oxygen
overabundance in galaxies in dense local environments (at $R_{0} = 1$
Mpc) is higher than that of galaxies in dense extended environments
(at $R_{0} = 5$ Mpc), i.e., the oxygen overabundance is more sensivite
to the local than to the extended environment density. 

Is the influence of the environment on the galactic chemical
abundances similar for galaxies of different masses?  We have divided
the sample of target galaxies $S030$ into two subsamples: the
``low-mass galaxies'' of stellar masses of $9.1 <$ log$M < 9.6$ and
``high-mass galaxies'' with masses of $9.6 <$ log$M < 10.1$.  The mean
oxygen overabundances $\Delta_{\rm OH}$ are determined for those
subsamples separately.  The panels in the right column of
Fig.~\ref{figure:n-doh-z03-v1000} show the mean oxygen overabundance
$\Delta_{\rm OH}$ of the high-mass galaxies (circles) and low-mass
galaxies (plus signs)  as a function of the number of the neighbour
galaxies.  The linear best fit to the data for the high-mass galaxies
is shown by the dashed line and that for low-mass galaxies by the
solid line.  Inspection of the panels in the right column of
Fig.~\ref{figure:n-doh-z03-v1000} suggests that the influence of the
environment on the abundances depends on galaxy mass, in the sense
that the low-mass galaxies show higher oxygen overabundance than the
high-mass galaxies.  

We use the median value of the overabundances for the 101 galaxies
with the densest environments at $R_{0} = 1$ Mpc. The median value of
the oxygen overabundance and the scatter in the overabundances is
$0.053 \pm 0.067$ dex for galaxies with masses of $9.1 <$ log$M < 9.6$
and $0.017 \pm 0.045$ dex for galaxies with masses of $9.6 <$ log$M <
10.1$.

Are the regions with the densest local environments (at $R_{0} = 1$
Mpc) also associated with high-density environments on larger scales?
Fig.~\ref{figure:n-n} shows the relation between the number of
neighbour galaxies at $R_{0} = 2$ Mpc and at $R_{0} = 1$ Mpc for
galaxies of $9.1 <$ log$M < 9.6$.  The circles represent data for 111
galaxies with more than 20 neighbour galaxies within $R_{0}$ = 1 Mpc.
The points denote data for 113 galaxies with more than 51 neighbour
galaxies within $R_{0} = 2$ Mpc.  Fig.~\ref{figure:n-n} shows that the
sample of regions with the densest local environments at $R_{0} = \sim
1$ Mpc  and the sample of regions with the densest environments  at
$R_{0} = 2$ Mpc overlap only partly.  The regions with the densest
local environments (at $R_{0} = 1$ Mpc) are not necessarily associated
with the highest-density regions on larger scales.  The regions with
high-density local environments can be found both in low-density and
high-density extended environments. 

So, there is an influence of the environment on the galactic
metallicities, i.e., the galaxies in dense environments tend to have
an overabundance  in oxygen. The influence of the environment on the
abundances depends on galaxy mass, in the sense that the low-mass
galaxies tend to show higher oxygen overabundances than the high-mass
galaxies.  However, the scatter in the oxygen overabundances is large
at any environmental density, i.e., galaxies with both enhanced and
reduced oxygen abundances can be found at any density of the
environment.  These conclusions can be checked using the nitrogen
abundances. It was noted above that the secondary nitrogen becomes
dominant at $12 +$ log(O/H) $\ga 8.2$, and, as consequence,  the
nitrogen abundance increases at a faster rate in comparison to the
oxygen abundance \citep{Pagel1997,Henry2000}. One can therefore expect
that the overabundance in nitrogen should have a larger amplitude than
that for oxygen. 

The panels in the left column of  Fig.~\ref{figure:n-dnh-z03-v1000}
show the mean nitrogen overabundances (points) and mean deviations of
the overabundances (plus signs) as a function of the number of
neighbour galaxies.  The panels in the middle column show the logarithm
of the number of target galaxies with a given number of neighbour
galaxies.  The panels in the right column
Fig.~\ref{figure:n-dnh-z03-v1000} show the mean nitrogen overabundance
$\Delta_{\rm NH}$ of the high-mass galaxies (circles) and low-mass
galaxies (plus signs) as a function of the number of neighbour
galaxies. 

Comparing Fig.~\ref{figure:n-doh-z03-v1000} and
Fig.~\ref{figure:n-dnh-z03-v1000} shows that the general behavior of
the oxygen and nitrogen overabundances is similar. Again there is a
tendency that the galaxies in dense environments are on average more
nitrogen-rich than other galaxies. The nitrogen overabundance is most
appreciable for low-mass galaxies in the densest environments and
decreases with increasing galaxy mass and decreasing environment
density.  The median values of the nitrogen overabundances and their
scatter for the 101 galaxies with the densest environments within
$R_{0} = 1$ Mpc are $0.109 \pm 0.172$ dex for galaxies with masses of
$9.1 <$ log$M < 9.6$ and $0.077 \pm 0.152$ dex for galaxies in the
mass range of $9.6 <$ log$M < 10.1$. As expected, the overabundances
in nitrogen are higher than the ones in oxygen.

The left panels of  Fig.~\ref{figure:n-dno-z03-v1000}
show the mean N/O excess (points) and mean deviations of
the excesses (plus signs) as a function of the number of
neighbour galaxies.  The middle panels show the logarithm
of the number of target galaxies with a given number of neighbour
galaxies.  The right panels of
Fig.~\ref{figure:n-dno-z03-v1000} show the mean excess 
$\Delta_{\rm NO}$ of the high-mass galaxies (circles) and low-mass
galaxies (plus signs) as a function of the number of neighbour
galaxies. 
Comparison of Fig.~\ref{figure:n-doh-z03-v1000}, Fig.~\ref{figure:n-dnh-z03-v1000}
and Fig.~\ref{figure:n-dno-z03-v1000} shows that the general behavior of
the N/O excesses is similar to that for oxygen and nitrogen overabundances.

Thus, the nitrogen abundances in our galaxies confirm the conclusions
reached from the analysis of the oxygen abundances. 

It was noted above that different numerical values of the peculiar
galactic velocities $V_{p}$ were adopted in different studies
\citep[e.g.,][]{Fuse2012,Argudo2015,Spector2016,Lacerna2016}. Above we
considered the case of  $V_{p} = 1000$ km s$^{-1}$. To evaluate the
influence of the adopted value of the peculiar velocities on the
results, we considered different values for this quantity.  The
results for the peculiar velocities $V_{p} = 100$ km s$^{-1}$ are
presented in Fig.~\ref{figure:n-doh-z03-v0100}.  The panels in the
left column of  Fig.~\ref{figure:n-doh-z03-v0100} show the mean oxygen
overabundances (points) and mean deviations of the overabundances
(plus signs) as a function of the number of the neighbouring galaxies
for the five values of $R_{0}$ given in each panel.  However, the bin
sizes differ from the ones used for $V_{p} = 1000$ km s$^{-1}$.  Here
the bin sizes in the number of neighbour galaxies are 1 for $R_{0} = 1$
Mpc,  3 for $R_{0} = 2$ Mpc,  5 for $R_{0} = 3$ Mpc,  8 for $R_{0} =
4$ Mpc,  and 10 for $R_{0} = 5$ Mpc.  The panels in the middle column
show the logarithm of the number of target galaxies having a given
number of neighbouring galaxies.  The panels in the right column of
Fig.~\ref{figure:n-doh-z03-v0100} show the mean oxygen overabundance
$\Delta_{\rm OH}$ of the high-mass (circles) and low-mass galaxies
(plus signs) as a function of the number of neighbour galaxies. 

The comparison between Fig.~\ref{figure:n-doh-z03-v1000} and
Fig.~\ref{figure:n-doh-z03-v0100} llustrates that the general behavior
of the oxygen overabundances is similar for $V_{p} = 1000$ km s$^{-1}$
and $V_{p} = 100$ km s$^{-1}$.  Again there is a tendency that the
galaxies in the dense environment are on average more oxygen-rich than
other galaxies. The overabundances decreases with increasing galaxy
mass and with decreasing environmental density. It is not surprising
that the number of neighbouring galaxies in the case of $V_{p} = 1000$
km s$^{-1}$ is larger than that for $V_{p} = 100$ km s$^{-1}$.

\subsection{Environmental influence on the abundances in galaxies 
of masses of $10^{10.5}$M$_{\sun}$ to $10^{11}$M$_{\sun}$}

\begin{figure}
\resizebox{1.00\hsize}{!}{\includegraphics[angle=000]{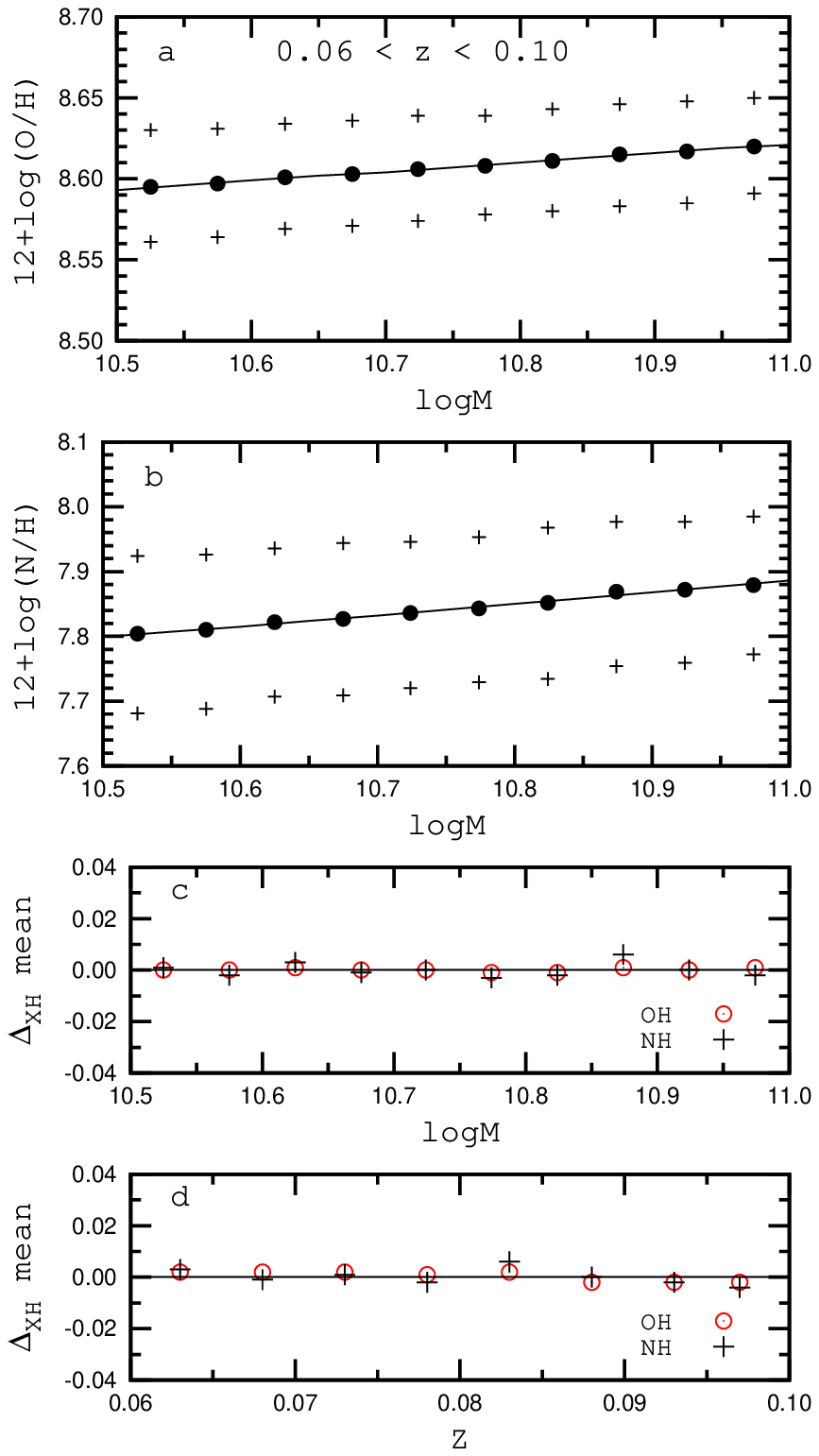}}
\caption{X/H -- $M$ relations for the sample of galaxies with masses of
$10.5 <$ log$M < 11$ and redhifts of $0.06 < z < 0.1$.  Panel $a.$ The
solid line shows the O/H $- M$ relation  The points represent the mean
oxygen abundances for galaxies in bins of 0.05 dex in stellar masses.
The plus signs indicate the mean deviations of the oxygen abundance
for the galaxies in the bins.  Panel $b.$ The same as Panel $a$ but
for nitrogen abundance.  Panel $c.$ The mean over(under)abundance of
oxygen (points) and nitrogen (plus signs) averaged for the galaxies in
bins of 0.05 dex in stellar mass as a function of galaxy mass.  Panel
$d.$ The mean over(under)abundance of oxygen (points) and nitrogen
(plus signs) averaged for galaxies in bins of 0.005 in redshift as a
function of redshift. 
}
\label{figure:xh-rel-z100}
\end{figure}

\begin{figure}
\resizebox{1.00\hsize}{!}{\includegraphics[angle=000]{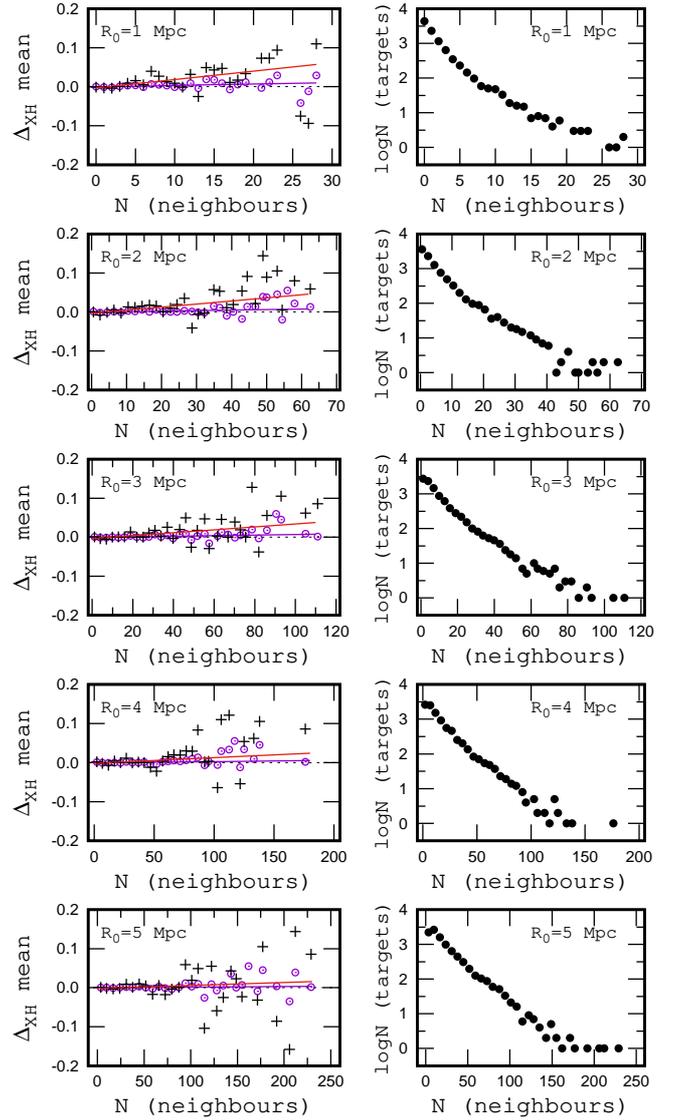}}
\caption{
Panels in the left column: Mean overabundance $d_{\rm XH}$ of galaxies
with redshifts of $0.6 \la z \la 0.1$ and stellar masses of $10.5 \la$
log$M \la 11$ as a function of number of neighbour galaxies within the
radius $R_{0}$ labeled in each panel.  The value of the peculiar
velocities of the neighbour galaxies is adopted to be 1000 km s$^{-1}$.
The mean overabundances in oxygen (circles) and nitrogen (plus signs)
are estimated for galaxies binned according to their number of
neighbour galaxies. The bins are 1 for $R_{0} = 1$ Mpc,  2 for $R_{0} =
2$ Mpc,  3 for $R_{0} = 3$ Mpc,  5 for $R_{0} = 4$ Mpc,  and 7 for
$R_{0} = 5$ Mpc.  The panels in the right column show the logarithm of
the number of target galaxies having a given number of neighbour
galaxies. 
}
\label{figure:n-dxh-z100-v1000}
\end{figure}

\begin{figure}
\resizebox{1.00\hsize}{!}{\includegraphics[angle=000]{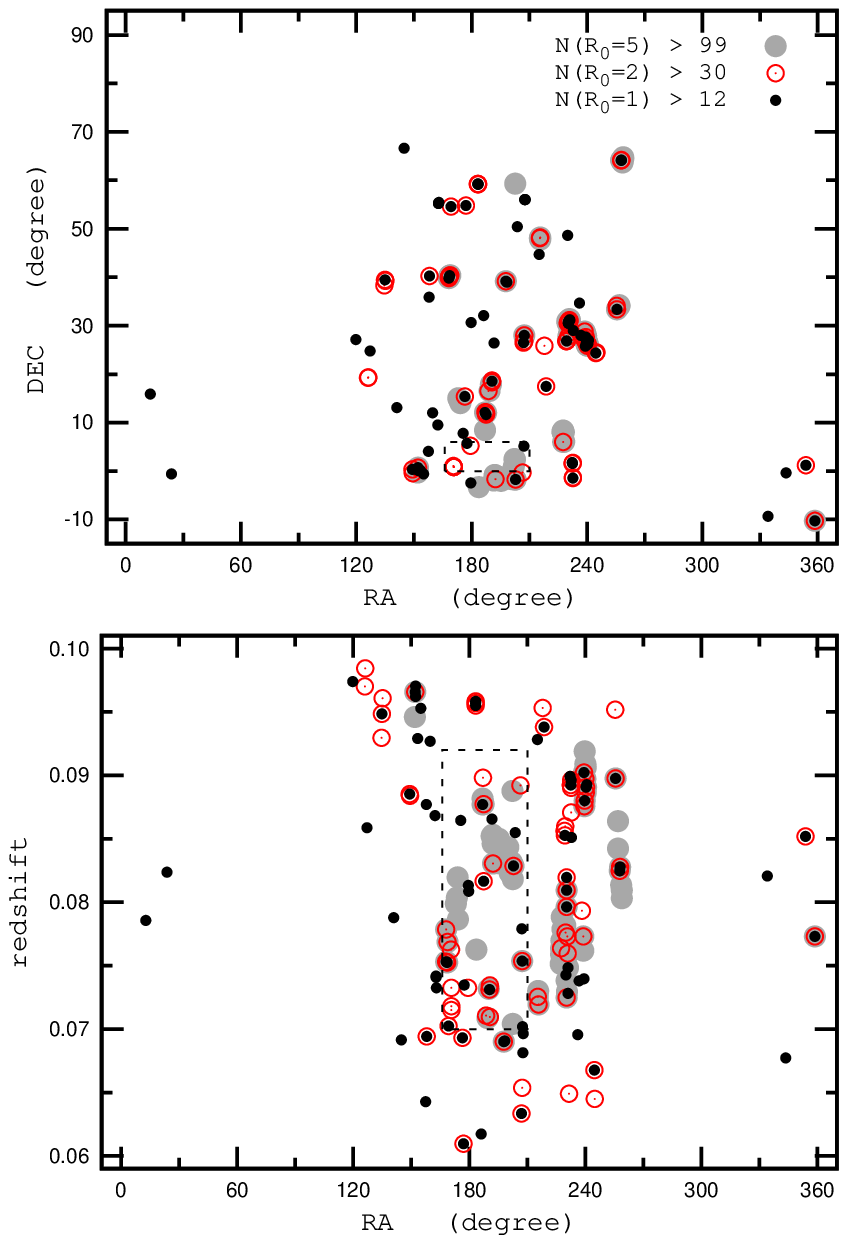}}
\caption{
Upper panel: Coordinate map of the galaxies from the sample $S100$
with the densest environment.  The points stand for galaxies for which
the number of neighbours at $R_{0} = 1$ Mpc is larger than 12, the dark
open circles depict galaxies for which the number of neighbours within
$R_{0} = 2$ Mpc is larger than 30, and the grey filled circles
indicate galaxies for which the number of neighbours within $R_{0} = 5$
Mpc is larger than 99.  The dashed line is roughly outlines the
position of the Sloan Great Wall.  The lower panel shows the location
of the galaxies from the upper panel in the RA vs. redshift plane. The
symbols are the same as in the upper panel.
}
\label{figure:ra-dec-z100}
\end{figure}

We showed in Section 2.1 that a realistic (complete) environmental
density of galaxies with masses of log$M \sim 10.5$ can be estimated
out to a redshift of $\sim 0.1$, see Fig.~\ref{figure:z-ngal-m}.  Here
we examine the influence of the environment on the abundances of
target galaxies with redshifts ranging from $z = 0.06$ to 0.10.  The
environmental density is estimated as the number of neighbouring
galaxies with masses higher than log$M \ga 10.5$. 

We consider the target galaxies in the stellar mass range from log$M =
10.5$ to 11.  We again neglect the evolutionary change  of the
chemical abundances, i.e., the dependence of the abundances on
redshift.  Fig.~\ref{figure:m-xh-z} suggests that the X/H -- $M$
relations for the galaxies within this stellar mass interval can be
reproduced by a simple expression.  Since we consider only massive
galaxies here we cannot correctly extract isolated galaxies.  Hence,
we determine the O/H -- $M$ relation for all galaxies of sample
$S100$. The resulting O/H -- $M$ relation is  
\begin{equation}
12 + \log({\rm O/H}) = 8.565 + 0.055 \,m + 0.00071 \, m^{2} ,
\label{equation:oh-z100}
\end{equation}
and the N/H -- $M$ relation is
\begin{equation}
12 + \log({\rm N/H}) = 7.725 + 0133 \,m + 0.027 \, m^{2} 
\label{equation:nh-z100}
\end{equation}
where $m =$ log$M - 10$. The galaxies with large deviations from the
O/H -- $M$ relation ($d_{O/H} > 0.15$ dex) or from the  N/H -- $M$
relation ($d_{N/H} > 0.4$ dex) are not used in deriving the final
relations and are excluded from further analysis (332 galaxies out of
9896). The mean deviations are $d_{OH}$ = 0.032 dex and  $d_{NH}$ =
0.118 dex based on 9564 galaxies.  Thus, our sample of target galaxies
with redshifts in the range of $0.10 \ga z \ga 0.06$ and stellar
masses of $11 \ga$ log$M \ga 10.5$ consists of 9564 galaxies and will
be referred to as sample $S100$.  This sample of galaxies includes the
galaxies from the very large overdense structure known as the Sloan
Great Wall \citep{Gott2005,Sheth2011}. 

The obtained O/H -- $M$ relation is shown in panel $a$ of
Fig.~\ref{figure:xh-rel-z100} by the solid line.  The filled circles
indicate oxygen abundances of galaxies averaged in bins of 0.05 dex in
stellar mass. The plus signs mark the mean deviations of the oxygen
abundance $d_{\rm OH}$ for those galaxies.  The N/H -- $M$ relation is
shown in panel $b$ of Fig.~\ref{figure:xh-rel-z100} by the solid line.
Again the filled circles denote nitrogen abundances averaged for
galaxies in bins of 0.05 dex in stellar mass, and the plus signs
indicate the mean deviations of the nitrogen abundance $d_{\rm NH}$.
Panel $c$ of Fig.~\ref{figure:xh-rel-z100} shows the mean
over(under)abundances of oxygen $\Delta_{\rm OH}$ (circles) and
nitrogen $\Delta_{\rm NH}$ (plus signs) for galaxies in stellar mass
bins of 0.05 dex as a function of galaxy mass, and panel $d$ shows the
mean over(under)abundances of oxygen (circles) and nitrogen (plus
signs) in redshift bins of 0.005 as a function of redshift. Panel $d$
of Fig.~\ref{figure:xh-rel-z100} demonstrates that there is no
appreciable trend of mean overabundance of oxygen and nitrogen with
redshift.  This confirms our assumption that the evolutionary change
of abundances (i.e., the dependence of the abundances on redshift) for
our sample of galaxies can be neglected. 

For each target galaxy we count the number of neighbouring galaxies in
the sample of environment galaxies $S_{EG}$.  The number of galaxies
in the region is determined using the neighbourhood criteria defined
earlier.  Again the five different values of $R_{0} = 1$ Mpc, 2 Mpc, 3
Mpc, 4 Mpc, and 5 Mpc are considered.  Only the case of a peculiar
velocity of $V_{p} = 1000$ km s$^{-1}$ is discussed. 

The panels in the left column of Fig.~\ref{figure:n-dxh-z100-v1000}
show the mean oxygen (circles) and nitrogen (plus signs) overabundance
as a function of the number of neighbouring galaxies within the volume
defined by the value of $R_{0}$  given in each panel.  The solid line
shows the best fit to the oxygen data and the dashed line to the
nitrogen data.  The panels in the right column show the logarithm of
the number of the target galaxies having the given number of
neighbouring galaxies.  The oxygen overabundances show a marginal (in
fact, negligibly small) correlation with the environmental density
while the trend with nitrogen overabundances is more noticeable.  The
median value of the overabundance and scatter in the overabundances
for the 101 galaxies with the densest environment at $R_{0} = 1$ Mpc
is $0.002 \pm 0.035$ dex for oxygen and $0.041 \pm 0.120$ dex for
nitrogen.  A comparison of Fig.~\ref{figure:n-doh-z03-v1000} and
Fig.~\ref{figure:n-dxh-z100-v1000} as well as of the median values of
the overabundances confirms the general trend that the overabundances
of galaxies in high-density environments decrease with increasing
galaxy mass. 

We also consider the case of peculiar velocities of $V_{p} = 100$ km
s$^{-1}$.  The general behavior of the oxygen and nitrogen
overabundances is similar for cases where  $V_{p}$ = 1000 km s$^{-1}$
and $V_{p}$ = 100 km s$^{-1}$ are used to count the number of
neighbouring galaxies.  Again the trend of overabundances with the
environmental density is insignificant for oxygen but more notable for
nitrogen overabundances. 

Our sample involves the galaxies from the large overdense structure
called the Sloan Great Wall. This permits us to take a closer look at
the distributions of the regions of the densest environments.  The
upper panel in Fig.~\ref{figure:ra-dec-z100} shows a sky map of
galaxies with the densest environments.  The black filled circles mark
the galaxies with the densest local environments, i.e., galaxies for
which the number of neighbours at $R_{0} = 1$ Mpc is larger than 12.
The dark (red) open circles denote galaxies for which the number of
neighbours at $R_{0} = 2$ Mpc is larger than 30.  The grey filled
circles indicate the galaxies with the densest extended environments,
i.e., galaxies for which the number of neighbours at $R_{5} =  5$ Mpc
is larger than 99. The lower panel shows the location of those
galaxies in the R.A.\ -- $z$ plane.  The Sloan Great Wall is roughly
outlined by the dashed line  stretching from R.A.\ $\sim 166 \degr$ to
R.A.\ $\sim 210 \degr$ and from Dec $\sim 0$ to Dec $\sim 6 \degr$  at
redshifts from $z \sim 0.07$ to $\sim 0.092$
\citep{Gott2005,Sheth2011}. 

Fig.~\ref{figure:ra-dec-z100} reveals that the high-density regions at
scales of 5 Mpc show a concentration (clustering) with several centres
including the Sloan Great Wall. The high-density regions at scales of
1 -- 2 Mpc show a more uniform distribution in space and do dot
exhibit a significant concentration to the Sloan Great Wall.  We again
found that the regions of the densest local environments (at $R_{0} =
1$ Mpc) are not nesessarily associated with the regions of the densest
environments at larger scales and the regions of the high-density
local environments can be found both in low-density and high-density
extended environments.

\section {Discussion}

The effects of the environment on the metal content of late-type
galaxies has been debated for a long time \citep[][among many
others]{Shields1991, Vilchez1995, Skillman1996,Pilyugin2002,
Mouhcine2007, Ellison2009, Hughes2013, Peng2014, Pustilnik2013,
Nicholls2014, Darvish2015, Kacprzak2015, Kreckel2015, Shimakawa2015,
Valentino2015}. 

These studies focus on comparisons between cluster and field galaxies.
For instance, \citet{Shields1991} determined the abundances in five
Virgo spirals of type Sc and found evidence that spiral galaxies in
the Virgo cluster have systematically higher interstellar abundances
than comparable field galaxies. 

\citet{Vilchez1995} examined the influence of the environment on the
evolution of dwarf galaxies. Subsamples of dwarf galaxies in nearby
voids, in the field of the local supercluster, and in the direction of
the core of the Virgo cluster were investigated. He concluded that
there is only marginal evidence in favor of a trend between the gas
metallicity and the density of the environment. Most of the galaxies
in his sample appear to follow to the general luminosity --
metallicity relation for dwarf galaxies.  Similar results were
obtained by \citet{Lee2007}, who found no pronounced dependence of the
luminosity--metallicity, metallicity--gas fraction, and metallicity --
tidal index diagrams on positive or negative tidal indices of dwarf
irregular galaxies in nearby galaxy groups of different environmental
densities. 

\citet{Mouhcine2007} investigated the dependence of gas-phase chemical
properties on stellar mass and environment using a sample of 37866
SDSS galaxies in the redshift range of $0.040 \la z \la 0.085$. They
found that galaxies show a marginal increase in chemical enrichment
level at a fixed stellar mass in denser environments. They concluded
that the evolution of star-forming galaxies is driven primarily by
their intinsic properties and is largely independent of their
environment over a large range of local galaxy densities.

\citet{Ellison2009} found that cluster galaxies in locally rich
environments have higher median metallicities by up to $\sim 0.05$ dex
than those in locally poor environments.  \citet{Hughes2013}
considered the relationship between stellar mass and metallicity for
260 nearby late-type galaxies in different environments, from isolated
galaxies to Virgo cluster members. They found that the mass --
metallicity relation is nearly independent of environment and
concluded that internal evolutionary processes, rather than
environment effects, play a key role in shaping the mass --
metallicity relation.  

\citet{Peng2014} found that there is a strong dependence on
environmental density for star-forming satellites (i.e., all galaxy
members of groups/clusters that are not located at the centre) and
that there is no correlation for galaxies located at the centre.
\citet{Darvish2015} studied the properties of a sample of 28
star-forming galaxies in a large filamentary structure at $z \sim
0.53$ (using Keck spectroscopic data) and compared them with a control
sample of 30 field galaxies.  They found that, on average,
star-forming galaxies in filaments are more metal-enriched ($\sim 0.1$
-- 0.15 dex).  

\citet{Shimakawa2015} considered the environmental dependence of
gaseous metallicities at redshifts of $z = 2.2$ and 2.5 and found that
the objects with masses below $10^{11}$M$_{\sun}$ tend to be more
chemically enriched than their field counterparts at $z  = 2.2$.
\citet{Valentino2015} investigated the environmental effect on the
metal enrichment of star-forming galaxies in a cluster at redshift $z
= 1.99$. They found that the cluster's star-forming galaxies in the
mass range of $10 \la$ log$M \la 11$ are up to 0.09 -- 0.25 dex more
metal-poor than their field counterparts, depending on the adopted
calibration for the abundance determinations.

\citet{Gupta2016} determined the abundances in galaxies of two 
galaxy clusters at redshifts $z$ $\sim$0.35. They found for one cluster 
that the galaxy abundance decreases as a function of projected 
cluster-centric distance. The galaxies of this cluster are shifted 
towards higher metallicity by a mean value of 0.2 dex compared to 
the local reference sample at a given mass. The galaxies of another 
cluster exhibit a bimodal radial abundance distribution, with 
one branch showing a high galaxy metallicity near the centre of the 
cluster and negative radial gradient and the other branch showing 
a low galaxy metallicity at the cluster centre and positive radial 
gradient. 
\citet{Gupta2016} suggested that the lower branch galaxies belong to a 
merging sub-sluster and/or are interloper galaxies.

Looking at it from another angle, voids are the most underdense
regions in the universe where galaxy evolution will have progressed
without any significant influence of the environment.  A number of
very low-metallicity galaxies have been revealed within void
environments \citep{Pustilnik2013}.  The nebular metallicities of two
dwarf galaxies in the Local Void were measured by
\citet{Nicholls2014}.  They find that the metallicities in both
galaxies are typical for galaxies of that size.  \citet{Kreckel2015}
measured gas-phase oxygen abundances and gas content for eight dwarf
galaxies located within the lowest density environments of seven void
galaxies and for a sample of isolated dwarf galaxies in average
density environments. They find no significant difference between
these void dwarf galaxies and the isolated dwarf galaxies.  They noted
that while the sample is too small to draw strong conclusions, their
result suggests that the chemical evolution of dwarf galaxies proceeds
independently of the large-scale environment.  \citet{Beygu2016}
studied the properties of 59 void galaxies and noted that their sample
consists of high metallicity galaxies as well as galaxies of low
metallicity.

We find for a large sample of galaxies that the oxygen and nitrogen
abundances tend to be slightly higher in galaxies located in volumes
with a high concentration of galaxies. The scatter in the abundances
is large at any concentration of galaxies (at any density of the
environment), i.e., galaxies with both enhanced and reduced abundances
can be found at any density of the environment.  Thus, our result
confirms the conclusions of the previous studies that environmental
effects do not play a key role in the chemical evolution of galaxies.

\section {Summary}

We examined the influence of the environment on the abundances of
late-type galaxies of masses of $10^{9.1}$ M$_{\sun}$ to $10^{11}$
M$_{\sun}$ with redsifts up to $z = 0.1$ from the Sloan Digital Sky
Survey (SDSS).  The oxygen and nitrogen abundances were estimated
through our recent calibrations and are compatible to the
$T_{e}$-based metallicity scale in  H\,{\sc ii} regions.  Galaxies
with reliable stellar masses and oxygen and nitrogen abundances were
extracted from the full sample. 

The obtained oxygen abundance -- galaxy mass (O/H -- $M$) and the
nitrogen abundance -- galaxy mass (N/H -- $M$) diagrams reproduce the
known evolution of galactic abundances with mass and redshift.  \\ The
oxygen and nitrogen abundances increase with increasing mass and with
decreasing redshift, and the X/H -- $M$ relations flattens out at the
massive end. \\ The rate of change of the chemical abundances in
galaxies at the present epoch decreases with increasing masses (the
downsizing effect).  \\ The rate of change of nitrogen abundance with
redshift and stellar mass is higher than that for oxygen. 

We studied the influence of the environmental density on the
overabundance of oxygen and nitrogen and investigated the deviation of
these abundances from the abundance -- galaxy mass relation.  The
environmental density was specified by the number of neighbouring
galaxies within a region in projected distance -- redshift difference
space.  Five different values of the projected distance were
considered: $R_{0} = 1$ Mpc, 2 Mpc, 3 Mpc, 4 Mpc, and 5 Mpc.  The
redshift differences take into account the change of redshift with
distance equal to the adopted $R_{0}$ and the peculiar velocities of
galaxies $V_{p}$. Three peculiar velocities were considered: $V_{p} =
1000$ km s$^{-1}$, 500 km s$^{-1}$, and 100 km s$^{-1}$.  

We find that the influence of the environment on the chemical
abundances of galaxies is most pronounced for galaxies of masses of
$10^{9.1}$ M$_{\sun}$ to $10^{9.6}$ M$_{\sun}$. The galaxies in the
densest environments can show an abundance excess larger than about
$\sim 0.05$ dex in oxygen (based on the median value for 101 galaxies
with the densest environments) and around $\sim 0.1$ dex in nitrogen
above the mean. The overabundance decreases with increasing galaxy
mass and with decreasing environmental density.

The overabundance in galaxies in dense local environments (at $R_{0} = 1$
Mpc) is higher than that of galaxies in dense extended environments
(at $R_{0} = 5$ Mpc), i.e., the oxygen overabundance is more sensivite
to the local than to the extended environment density.

Since only a small fraction of galaxies are in high-density
environments then they do not have a significant influence on the
general X/H -- $M$ relation.  The metallicity -- mass relations for
isolated galaxies and galaxies having neighbours are similar to each
other.  The mean shift of non-isolated galaxies around the metallicity
-- mass relation traced by the isolated galaxies is less than $\sim
0.01$ dex for oxygen and less than $\sim 0.02$ dex for nitrogen.  The
scatter in the overabundances is large at any environmental density,
i.e., galaxies with both enhanced and reduced abundances can be found
at any density of the environment. This suggests that environmental
effects do not play a key role in the chemical evolution of late-type
galaxies.  This conclusion has been reached in a number of previous
studies, and our results confirm that conclusion.

\section*{Acknowledgements}

We are grateful to the referee for his/her constructive comments. \\
L.S.P.,\ I.A.Z.\ and E.K.G.\ acknowledge support in the framework of
Sonderforschungsbereich SFB 881 on "The Milky Way System" (especially
subproject A5), which is funded by the German Research Foundation
(DFG). \\ 
L.S.P.\ and I.A.Z.acknowledge support of the Volkswagen Foundation under
the Trilateral Partnerships grant No. 90411. \\
L.S.P.\ and I.A.Z. thank for the hospitality of the Astronomisches
Rechen-Institut at Heidelberg University  where part of this
investigation was carried out. \\
This work was partly funded by the subsidy allocated to Kazan Federal
University for the state assignment in the sphere of scientific
activities (L.S.P.).  \\ 
Funding for the SDSS and SDSS-II has been provided by the Alfred P. Sloan Foundation, 
the Participating Institutions, the National Science Foundation, the U.S. Department of Energy,
the National Aeronautics and Space Administration, the Japanese Monbukagakusho, the Max Planck Society, 
and the Higher Education Funding Council for England. The SDSS Web Site is http://www.sdss.org/.

\end{document}